\documentclass[aps,pre,groupedaddress,longbibliography]{revtex4-1}

\usepackage{graphicx}
\usepackage[normalem]{ulem}
\usepackage{color}
\usepackage{soul}

\newcommand*{\Ray}{{\rm Ra}}
\newcommand*{\Ek}{{\rm E}}

\newcommand{\Ro}{{\rm Ro}}
\newcommand{\Ree}{{\rm Re}}
\newcommand{\uu}{{\rm u}}

\renewcommand{\epsilon}{\varepsilon}

\newcommand{\bigD}{{\rm D}}

\def\bfu{\mbox{\bf u}}
\def\bfe{\mbox{\boldmath $\hat{e}$}}
\def\bfnabla{\mbox{\boldmath $\nabla$}}
\def\bfomega{\mbox{\boldmath $\omega$}}
\def\bfg{\mbox{\bf g}}

\def\stokes{\mbox{$\nabla_{\star}^{2}$}}

\begin{document}



\title{On the Formation of Eyes in Large-scale
Cyclonic Vortices}



\author{L. Oruba} 
\thanks{Now at LATMOS (IPSL/UPMC/CNRS).}
\affiliation{Physics Department, Ecole Normale
  Sup\'{e}rieure,\\ 24 rue Lhomond, 75005 Paris, France.} 
\author{ P.~A. Davidson} 
\affiliation{Engineering Department, University of Cambridge,
\\ Trumpington Street, Cambridge CB2 1PZ, UK.}
\author{E. Dormy} 
\affiliation{Department of Mathematics \& Applications, CNRS UMR 8553, 
Ecole Normale Sup\'{e}rieure,\\ 45 rue d'Ulm, 75005 Paris, France.}


\date{\today}

\begin{abstract}

We present numerical simulations of steady, laminar, axisymmetric convection 
of a Boussinesq fluid in a shallow, rotating, cylindrical domain. 
The flow is driven by an imposed vertical heat flux and shaped by the 
background rotation of the domain. The geometry is inspired by that of 
tropical cyclones and the global flow pattern consists of a shallow, 
swirling vortex combined with a poloidal flow in the $r-z$ plane which 
is predominantly inward near the bottom boundary and outward along the 
upper surface. Our numerical experiments confirm that, as suggested 
by \cite{ODD17}, an eye forms at the centre of 
the vortex which is reminiscent of that seen in a tropical cyclone 
and is characterised by a local reversal in the direction of the 
poloidal flow. We establish scaling laws for the flow and map out 
the conditions under which an eye will, or will not, form. We show 
that, to leading order, the velocity scales with 
$V=(\alpha g \beta)^{1/2}H$, 
where $g$ is gravity, $\alpha$ the expansion coefficient,  
$\beta$ the background temperature gradient, and 
$H$ is the depth of the domain. We also show that the two most important 
parameters controlling the flow are $\Ree={V H}/{\nu}$ and 
$\Ro={V}/\left(\Omega H \right)$, where $\Omega$ is the background 
rotation rate and $\nu$ the viscosity. The Prandtl number and aspect ratio 
also play an important, if secondary, role. Finally, and most importantly, 
we establish the criteria required for eye formation. 
These consist 
of a lower bound on $\Ree$, upper and lower bounds on $\Ro$, and an upper 
bound on Ekman number. 

\end{abstract}

\pacs{}

\maketitle

\section{Introduction}

A well-documented and intriguing feature of atmospheric vortices, such as 
tropical cyclones and dust-devils, is that they often develop an eye, defined 
as a region of reversed, downward flow in and around the axis of the vortex 
\citep[see][and references therein]{Lugt}. In the case of tropical cyclones, such 
an eye is readily identified in satellite images by the absence of cloud cover. 
Despite their common appearance, there is still little agreement as to the 
mechanisms of eye formation \citep[]{Pearce, Smith, Pearce2}, and 
indeed it is not even clear that the same 
basic mechanisms are responsible in 
different classes of atmospheric vortices \citep{Rotunno14}. In the absence of such a 
fundamental understanding, one cannot reliably predict when eyes should, 
or should not, form. 

\begin{figure}
\centerline{
\includegraphics[width=0.6\textwidth]{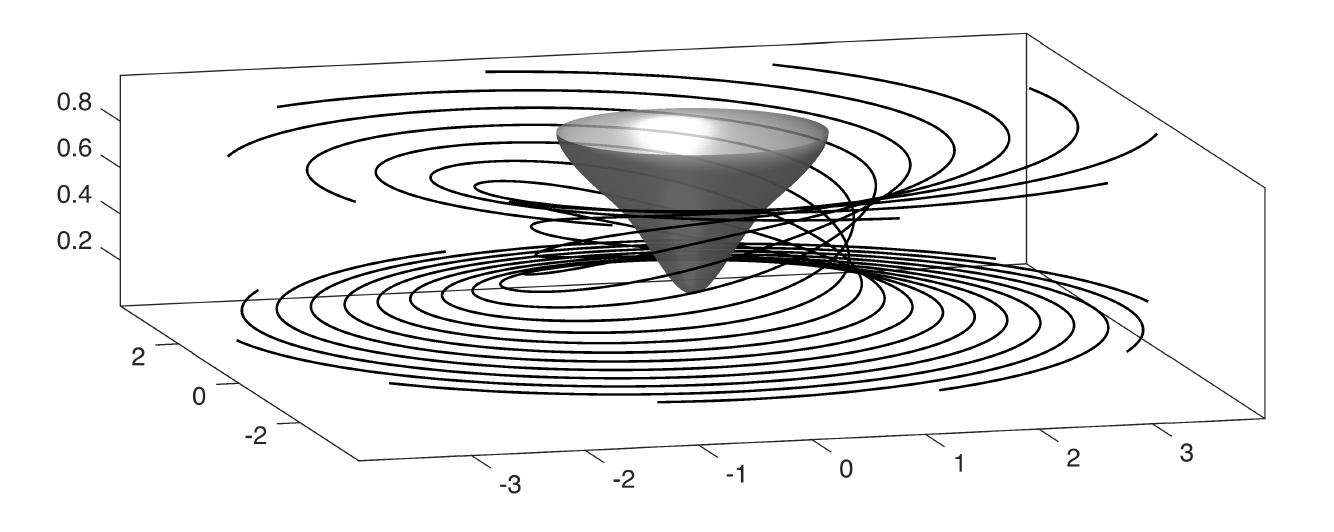}
}
\caption{Streamlines highlighting the cyclonic vortex in the center of our convection
    cell. The conical eyewall is represented in shaded gray. Parameters: 
$\epsilon=0.1$, $\Pr = 0.1$, $\Ek = 0.1$ and $\Ray = 2 \times 10^4$.}
\label{fig1}
\end{figure}

Recently, however, Oruba {\it et al} \cite{ODD17}, (hereafter denoted ODD17) identified one 
mechanism of eye formation in the context of a simple model problem. Inspired 
by the geometry of tropical cyclones, they considered convection in a shallow, 
rotating, cylindrical domain of low aspect ratio. In particular, they 
investigated the simplest physical system that can support an eye in such a 
geometry, which is the steady, laminar, axisymmetric convection of a Boussinesq 
fluid. 
Such a simple system is free from the complexities which hamper our 
understanding of real atmospheric vortices, such as turbulence, stable 
stratification, ill-defined boundary conditions, latent heat release from 
moist convection, and transient evolution. This allowed the mechanism of eye 
formation to be unambiguously identified, at least for the model system 
considered. It turns out that the eye in such cases is a passive response 
to the formation of an eyewall, a thin conical annulus of upward moving fluid 
which forms near the axis and separates the eye from the rest of the vortex 
(see Figure~\ref{fig1}).
Such eyewalls are characterised by a particularly intense level of negative 
azimuthal (horizontal) vorticity, and ODD17 showed that the eye, which is also 
characterised by a region of negative azimuthal vorticity, receives its 
vorticity by slow, cross-stream diffusion from the eyewall. Since the main 
body of the vortex has positive azimuthal vorticity, it is natural to ask 
where the intense, negative azimuthal vorticity of the eyewall comes from, 
and ODD17 established that the eyewall vorticity has its origins in the 
boundary layer on the bottom surface. 

Perhaps it is worth taking a moment to describe the model system of 
ODD17, if only because we shall adopt the same system here. It consists of 
a rotating, cylindrical domain of low aspect ratio in which the lower surface 
is a no-slip boundary and the upper surface is stress free. 
The motion is driven by a prescribed vertical heat flux through 
the lower boundary and, in a frame of reference rotating with the lower 
boundary, the flow is organised and shaped by the Coriolis force. Crucially, 
this Coriolis force induces positive excess swirl in the fluid adjacent to 
the lower boundary, which in turn sets up an Ekman-like boundary layer on the 
lower surface. This boundary layer then drives flow inward towards the axis, 
and so the primary motion in the vertical plane is radially inward near the 
lower boundary and outward at the upper surface. As the fluid spirals inward, 
it tries to conserve its angular momentum, and this results in a region of 
particularly intense swirl near the axis. 

In the force balance for the bulk of the vortex it was found that the 
buoyancy, Coriolis and inertial forces are of similar magnitudes, with a local 
Rossby number of order unity. However, near the eyewall the intense swirl 
means that the local Rossby number is large, with the buoyancy and Coriolis forces 
almost completely irrelevant by comparison with inertia. So, surrounding the 
eyewall there exists a conventional converging, swirling boundary layer, which 
separates before reaching the axis, carrying its intense azimuthal vorticity 
up into the bulk of the flow. The resulting free shear layer then constitutes 
the eyewall, which in turn gives rise to an eye.

For the limited range of parameters considered in ODD17, the requirement for  an 
eye to form is that the Reynolds number based on the peak inflow velocity must exceed 
$\Ree \sim 37$. By contrast, at lower values of Re the flow is relatively diffusive, 
and so the negative azimuthal vorticity in the lower boundary layer cannot 
be advected upward to form an eyewall; hence the absence of an eye. 
To the best of our knowledge, 
this is the first attempt to establish a simple criterion for eye formation. 
However, ODD17 considered only a relatively small range of parameters, 
keeping the aspect ratio and Ekman number fixed and varying the Rayleigh 
number by a factor of only $30$. 
Here we revisit the entire problem and consider a much wider range of parameters. 
In particular, we present the results of a suite of over 
150 numerical simulations 
in which the Rayleigh number, Ekman number and aspect ratio are all varied. The 
analysis of this suite of simulations shows that the conditions required for eye 
formation are more subtle than those suggested in ODD17.

\section{A Model Problem and Key Dimensionless Groups}

Our model problem is the same as that in \cite{ODD17}. It consists of the steady, 
laminar flow of a Boussinesq fluid in a closed, rotating cylinder of height $H$ 
and radius $R$, with aspect ratio $\epsilon=H/R \ll 1$. We adopt cylindrical polar 
coordinates, $(r, \phi, z)$, with the upper and lower boundaries at $z=H$ and 
$z=0$. 
The motion is maintained by buoyancy with a 
prescribed heat flux between the two horizontal boundaries. 
The surfaces at $z=0$ and $r=R$ are no-slip boundaries, while the upper 
surface is taken to be stress free. 
This choice of boundary conditions is essential to our model
as the vorticity generation in the bottom boundary layer is
essential in the eyewall formation. It does not necessarily imply that
counter-vortices  (associated with a downward flow near the axis) are not 
possible under different
configurations, such as for example a stress-free bottom boundary. However
such structures would not feature a sharp eyewall as in the present model.

  The choice of fixed heat flux boundary condition is motivated
  by our intention to model an elongated vortex: we want to drive a
  large-scale convective cell in an elongated domain. It is well known
  \citep[e.g.][]{Turner} that imposed flux boundary conditions will cause the
  convective cell to extend horizontally and fill the entire domain. This choice of boundary 
  conditions is also the natural
  choice to model intense atmospheric vortices over the ocean, the main
  source of energy being the flux of water vapour from the ocean.

In the absence of 
convection there is an imposed, uniform temperature gradient of 
${\rm d}T_0/{\rm d}z=-\beta$, and we write the 
temperature distribution in the presence of convection as 
$T=T_0(z) + \vartheta$. The governing equation 
for the temperature disturbance is then
\begin{equation}
\frac{\bigD \vartheta}{\bigD t}=\kappa \, \bfnabla^2 \vartheta + \beta u_z \,,
\label{eqtheta} 
\end{equation}
where $\kappa$ is the thermal diffusivity and $u_z$ the vertical velocity. 
We impose $\partial \vartheta/\partial z$ at $z=0$ and $z=H$ in order to 
maintain a constant axial heat flux, and the outer radial boundary is taken 
to be thermally insulating. 

Let ${\bf \Omega}$ be the background rotation rate, and $\nu$, $\alpha$ and 
$\rho_0$ be the kinematic viscosity, expansion coefficient and mean density of the 
fluid. In a frame of reference which rotates with the boundaries $z=0$ and 
$r=R$, the governing equation of motion is then 

\begin{equation}
\frac{\bigD {\bf u}}{\bigD t} = 
\, - {\bfnabla}\left(p/\rho_0\right) 
\, + 2 \, {\bf u} \times {\bf \Omega} 
\, + \nu \bfnabla^2 {\bf u}
\, - \alpha \vartheta {\bf g} \,,
\label{equ}
\end{equation}	
where ${\bf u}$ is the solenoidal velocity field in the rotating frame, 
$p$ the departure from a hydrostatic pressure distribution, and 
$-\alpha \vartheta {\bf g}$ the buoyancy force per unit mass. 
The associated vorticity equation is 
\begin{equation}
\frac{\bigD {\bfomega}}{\bigD t} =  {\bfomega} \cdot {\bfnabla} {\bf u} 
\, + 2 \, {\bf \Omega}  \cdot {\bfnabla} {\bf u} 
\, + \nu \bfnabla^2 {\bfomega}
\, + \alpha {\bf g} \times \bfnabla \vartheta  \,,
\label{eqomega}
\end{equation}	
where ${\bfomega}={\bfnabla} \times {\bf u}$.

Since we restrict ourselves to axisymmetric velocity fields it is 
convenient to decompose ${\bf u}$ into poloidal, 
${\bf u}_p=(u_r, 0, u_z)$, and azimuthal, 
${\bf u}_\phi=(0, \Gamma/r, 0)$, components, in which $\bfnabla \cdot {\bfu}_p=0$ and 
$\Gamma=ru_\phi$ is the angular momentum density in the rotating frame. 
The azimuthal component of (\ref{equ}) and (\ref{eqomega}) 
then becomes evolution equations for $\Gamma$ and 
$\omega_\phi=\left(\nabla \times {\bfu}_p\right)\cdot \bfe_\phi$,
\begin{eqnarray}
&\displaystyle \frac{\bigD}{\bigD t}& \left(\Gamma + \Omega r^2\right) =
\, \nu \stokes {\Gamma} \label{eqGamma}\\
&\displaystyle \frac{\bigD}{\bigD t} &
\left(\frac{\omega_\phi}{r}\right) = \frac{\partial}{\partial
  z}\left(\frac{\Gamma^2}{r^4}\right) + \frac{2\, \Omega}{r} \frac{\partial
  u_\phi}{\partial z}  -\frac{\alpha g}{r}\,   \frac{\partial
  \vartheta}{\partial r} + \frac{\nu}{r^2}\stokes\left(r \omega_\phi \right)
\,,
\label{eqomegaphi}
\end{eqnarray}
where $\stokes$ is the Stokes operator,
\begin{equation}
\stokes=r \frac{\partial }{\partial r}\left(\frac{1}{r}\frac{\partial }{\partial r}
\right) + \frac{\partial^2 }{\partial z^2} \,.
\end{equation}
\citep[See, for example,][for a derivation of equations 
(\ref{eqGamma}) and (\ref{eqomegaphi})]{Davidson}. The Stokes stream-function, 
defined by $\bfu_p= \bfnabla \times 
\left[\left(\psi/r\right) \bfe_\phi\right]$, can be determined from $\omega_\phi$ 
by inverting the Poisson equation $r\omega_\phi=-\stokes \psi$. It follows that 
the two scalar fields $\Gamma$ and $\omega_\phi$ uniquely determine 
the instantaneous velocity distribution, and so the governing equations for our 
model system are (\ref{eqtheta}), (\ref{eqGamma}) and (\ref{eqomegaphi}).

The dimensionless control parameters normally used to investigate the stability 
of this kind of rotating convection are 
\begin{equation}
\epsilon=\frac{H}{R} \, ,
\qquad 
\Pr=\frac{\nu}{\kappa} \, ,
\qquad 
\Ek=\frac{\nu}{\Omega H^2} \, , 
\qquad 
\Ray=\frac{\alpha g \beta H^4}{\nu \kappa}\,,
\label{Parameters1}
\end{equation}	
where $\Pr$ is the Prandtl number, $\Ek$ the Ekman number and $\Ray$ the Rayleigh 
number. However, since we are looking at fully-developed flow, rather than the 
stability of a static equilibrium, we shall find it convenient to
work with an alternative set of dimensionless parameters. Let us introduce the 
velocity scale  $V=(\alpha g \beta)^{1/2}H$, which will turn out to be characteristic 
of the actual fluid velocity. Then an alternative, if equivalent, 
set of dimensionless control parameters is
\begin{equation}
\epsilon=\frac{H}{R} \, ,
\qquad 
\Pr=\frac{\nu}{\kappa} \, ,
\qquad 
\Ree=\frac{V H}{\nu} \, ,
\qquad 
\Ro=\frac{V}{\Omega H}\,,
\label{Parameters2}
\end{equation}	
where $\Ree$ and $\Ro$ are characteristic Reynolds and Rossby numbers. One potential 
advantage of (\ref{Parameters2}) over (\ref{Parameters1}) is that, if we are allowed 
to take $V$ as truly representative of fluid velocity, then $\Ree$ and $\Ro$ have a 
simple physical interpretation in terms of the relative dynamical balance
in (\ref{equ}). 
The dimensionless control parameters (\ref{Parameters2}) indeed naturally enter the 
non-dimensional form of equations (\ref{eqtheta}) and (\ref{equ}) 
using $H$, $V$ and $\beta H$ as units of length, speed and temperature, which 
provides
\begin{eqnarray}
\frac{\bigD \vartheta^\star}{\bigD t^\star}&=&\Ree^{-1} \,{\Pr}^{-1} 
\, \bfnabla^2 \vartheta^\star + u_z^\star \\
\frac{\bigD {\bf u^\star}}{\bigD t^\star} &=& 
\, - {\bfnabla} \pi 
\, + 2 \, \Ro^{-1} \,{\bf u^\star} \times {\bfe_z} 
\, + \Ree^{-1} \, \bfnabla^2 {\bf u^\star}
\, + \vartheta^\star {\bfe_z} \,,
\label{eqadim} 
\end{eqnarray}
where a $^\star$ denotes dimensionless quantities.
Moreover, ODD17 have already noted the importance of $\Ree$ as a control 
parameter for the appearance of an eye. Of course, it is easy to go from 
(\ref{Parameters1}) to (\ref{Parameters2}), with  
$\Ro=\Ray^{1/2}\Ek\Pr^{-1/2}$ and $\Ree=\Ray^{1/2}\Pr^{-1/2}$. 

The eyewall tends to be confined to the region $r<H$ and, as noted above, the 
dynamics in the vicinity of the eyewall tends to be quite different to be that in 
the bulk of the vortex. In particular, although the Coriolis and buoyancy forces 
are of the same order of magnitude as inertia in the bulk, they are negligible near 
the eyewall where inertia is particularly high. Consequently, for diagnostic 
purposes, we shall find it convenient to introduce the following local definitions of 
$\Ree$ and $\Ro$. Let $u_{\phi,m}$  be the maximum azimuthal velocity on the surface 
$r = H$, and $u_{r,\delta}$ be the magnitude of the radial velocity at location 
$\left(r=H, z=\delta\right)$, where $z=\delta$ is the upper edge of the bottom 
boundary layer, defined at a given radius as the point where negative 
azimuthal vorticity $\omega_\phi$ in the boundary layer becomes positive.  
We then define local values of $\Ree$ and $\Ro$ in the vicinity of the eyewall 
as $\Ree_r=u_{r,\delta} H/\nu$, $\Ree_\phi=u_{\phi,m} H/\nu$, 
$\Ro_r=u_{r,\delta}/(\Omega H)$ and $\Ro_\phi=u_{\phi,m}/(\Omega H)$. More generally, 
we introduce local values of $\Ro_r(r)$ and $\Ro_\phi(r)$ for any radius, based 
on the local values of $u_{r,\delta}(r)$ and $u_{\phi,m}(r)$.

The numerical values of the dimensionless control parameters used in our suite 
of numerical simulations are tabulated in the Appendix, along with the corresponding 
values of $\Ro$, $\Ree$, $\Ro_r$, $\Ree_r$ and the magnitude of 
the maximum downward velocity on the axis, $| \uu_z |^{\rm max}_{\rm r=0}$. 
The dimensionless parameters listed in (\ref{Parameters1}) are restricted to 
the ranges $0.1<\epsilon<0.3$, $0.1<\Pr<1$, $0.07<\Ek<0.4$ 
and $10^3<\Ray<4.5 \times 10^4$. These 
correspond to values of $\Ree$, $\Ro$ and $\Ree_r$ of 
$45<\Ree<616$, 
$4.5<\Ro<124$ and
$6<\Ree_r<188$. 
A zero entry for $| \uu_z |^{\rm max}_{\rm r=0}$ in the table 
indicates that no eye formed in that simulation, while a non-zero value provides 
a measure of the strength of the eye.

There are 157 simulations in total. Each numerical experiment 
comprises an initial value problem which is run until a steady state is reached. 
We solve equations (\ref{eqtheta}), (\ref{eqGamma}) and (\ref{eqomegaphi}) using 
second-order finite differences with an implicit second-order backward 
differentiation in time. The grid resolution is $1000$ radial $\times 500$ 
axial cells and grid resolution studies were performed to ensure convergence.

\begin{figure}
\centerline{
\includegraphics[width=1\textwidth]{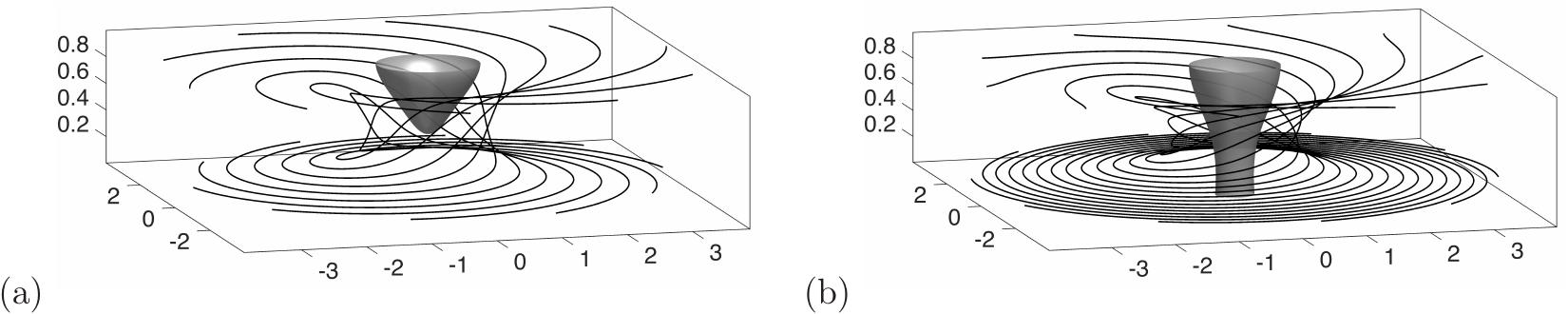}
}
\caption{Similar representation as Figure~\ref{fig1} for
different choices of parameters: (a) 
$\epsilon=0.1$, $\Pr = 0.1$, $\Ek = 0.15$ and $\Ray = 1 \times 10^4 \,,$
(b)
$\epsilon=0.1$, $\Pr = 1$, $\Ek = 0.1$ and $\Ray = 4.5 \times 10^4 \, .$}
\label{fig2}
\end{figure}

The strength and shape of the eye depends on the parameter regime, see table 1, and figure~\ref{fig2}.

\section{General Flow Structure and Scaling Laws}

As a prelude to our discussion of the conditions under which eyes form, it is useful 
to consider the general structure of the flow and the scaling laws for the velocity 
field. In order to illustrate some of the more general features of the flow, let us 
start by considering the specific (though typical) case in which the control 
parameters are $\epsilon=0.1$, $\Pr = 0.1$, $\Ek = 0.1$ and $\Ray = 2 \times 10^4$, 
or equivalently  $\Ree = 447$ and $\Ro = 44.7$. The Reynolds number and Rossby 
number in the vicinity of the eyewall are $\Ree_r=176$ and $\Ro_\phi=30$. 

\begin{figure}
\centerline{
\includegraphics[width=0.8\textwidth]{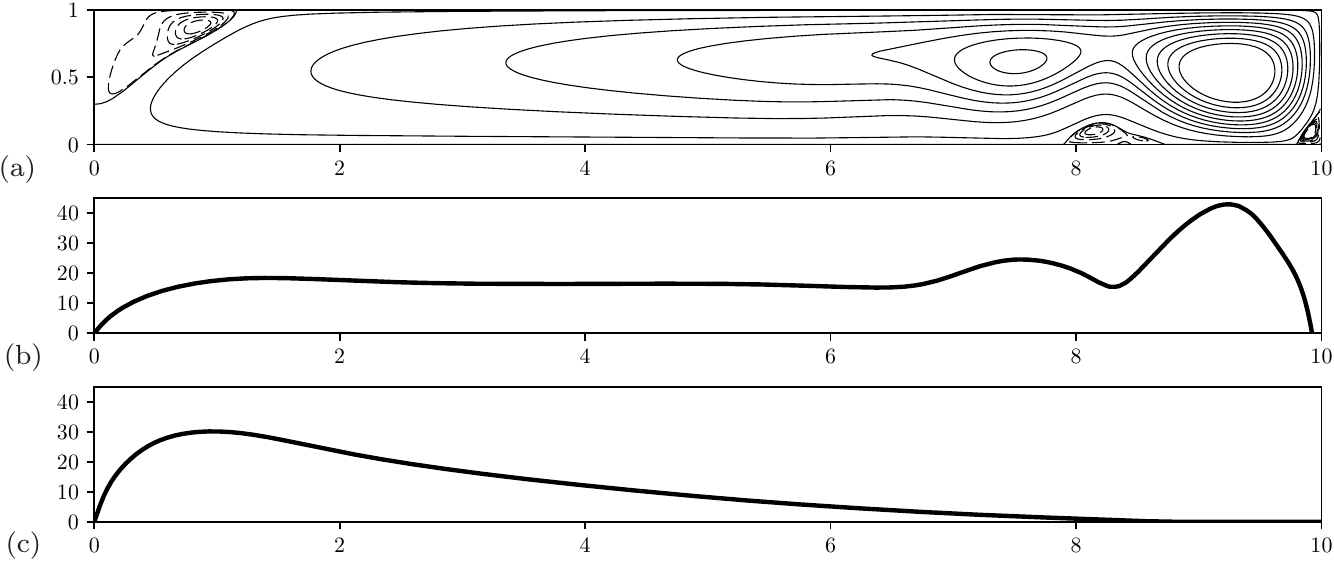}
}
\caption{(a) The Stokes stream-function. (b) The radial variations of 
$\Ro_r(r)$ and (c) $\Ro_\phi(r)$. Parameters: 
$\epsilon=0.1$, $\Pr = 0.1$, $\Ek = 0.1$ and $\Ray = 2 \times 10^4$.}
\label{fig3}
\end{figure}

The Stokes stream-function and radial variations of $\Ro_r(r)$ and 
$\Ro_\phi(r)$ for this case are shown in Figure~\ref{fig3}, and it is evident 
that an eye has formed near the axis. 
Note that $\Ro_\phi(r)$, and hence $u_\phi$, rises rapidly 
as we approach the eyewall, which is a consequence of approximate angular momentum 
conservation in the incoming flow. The local value of $\Ro_\phi$ near the eye is 
therefore large and background rotation has no direct influence on the flow in this 
region. Note also that $u_{\phi,m}$ is smaller than $u_{r,\delta}$ in the bulk of the 
vortex, but that $u_{\phi,m}$ exceeds $u_{r,\delta}$ near the eyewall. 

\begin{figure}
\centerline{
\includegraphics[width=0.8\textwidth]{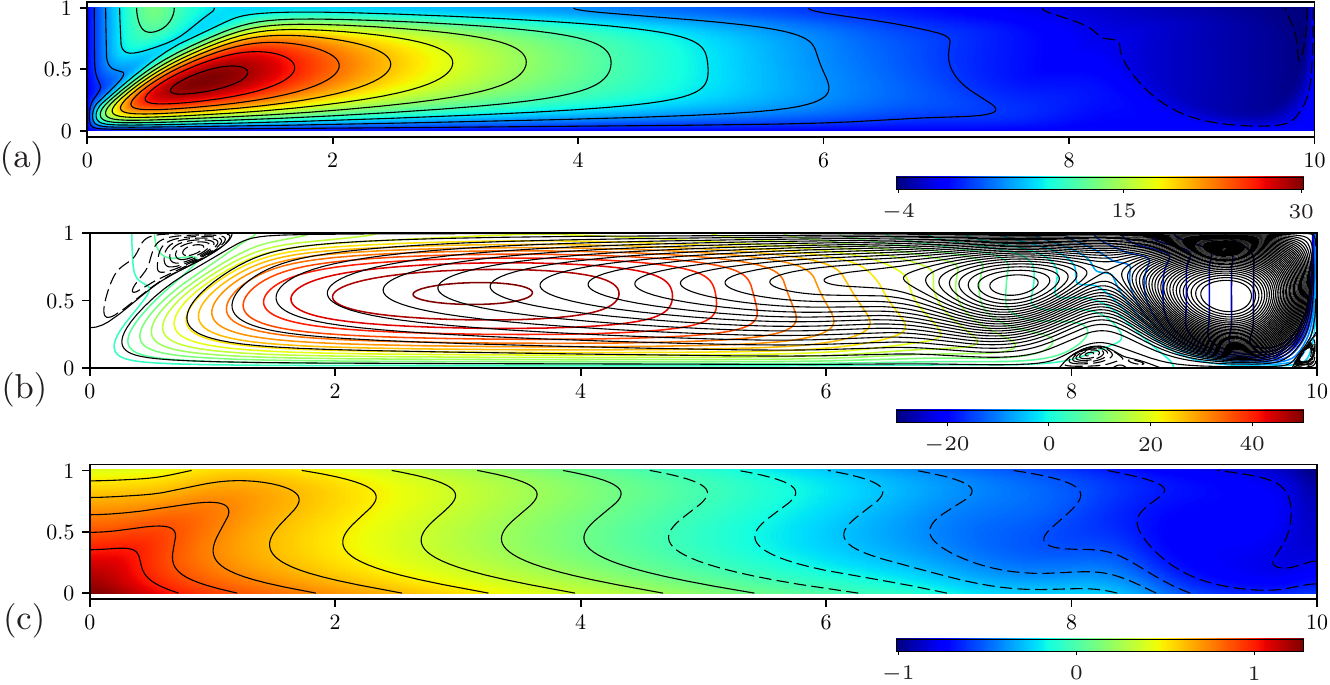}
}
\caption{Colour maps of: (a) the azimuthal velocity, $u_\phi$, (b) the angular 
momentum, $\Gamma$, superimposed on the stream-function, and (c) 
the total temperature $T=T_0(z)+\vartheta$. 
Parameters: $\epsilon=0.1$, $\Pr = 0.1$, $\Ek = 0.1$ and $\Ray = 2 \times 10^4$.}
\label{fig4}
\end{figure}

Figure~\ref{fig4} shows the corresponding distributions of azimuthal velocity, 
$u_\phi$, angular momentum, $\Gamma$, and total temperature, $T=T_0(z)+\vartheta$. 
The intensification of $u_\phi$ by the inward advection of angular momentum is 
evident in Figure~\ref{fig4}(a), while ~\ref{fig4}(b) shows that, in the region 
immediately to the right of the eye, the contours of constant angular 
momentum are roughly aligned with the 
streamlines, indicative of $\bigD {\Gamma}/{\bigD t} \simeq 0$. 
This is to be expected from (\ref{eqGamma}), given that the 
background rotation is locally weak and diffusion is largely restricted to the 
boundary layer and the eyewall. This figure also shows a substantial region of 
negative $u_\phi$ (anti-cyclonic rotation) at large radius, 
something that is also noted in ODD17 and is observed in tropical cyclones. 
From Figure~\ref{fig4}(c) we see that the 
poloidal flow sweeps hot fluid upward near the axis and cold fluid downward and 
inwards at $r = R$. The resulting negative radial gradient in temperature drives the 
main poloidal vortex, ensuring that it has positive azimuthal vorticity in accordance
 with equation (\ref{eqomegaphi}). 

\begin{figure}
\centerline{
\includegraphics[width=0.82\textwidth]{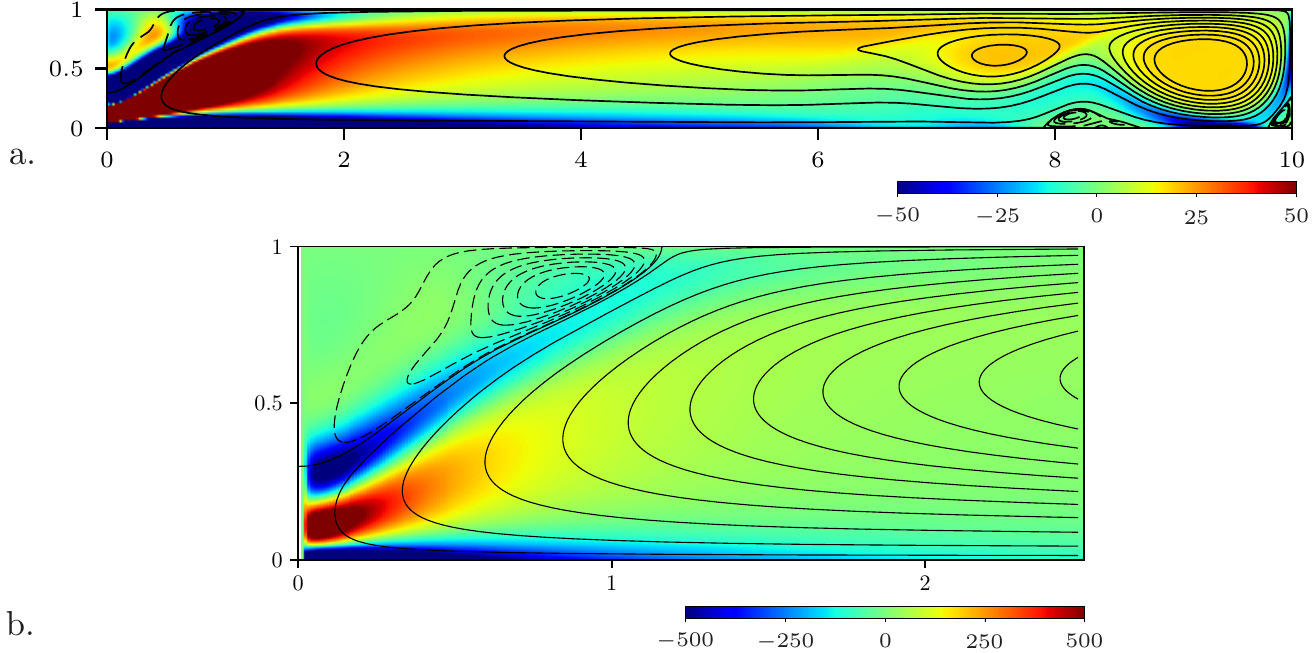}
}
\caption{Colour map of $\omega_\phi/r$ 
superimposed on the streamlines. (a) Full flow field. 
(b) Flow in the inner quarter of the domain.  
Parameters: $\epsilon=0.1$, $\Pr = 0.1$, $\Ek = 0.1$ and $\Ray = 2 \times 10^4$.}
\label{fig5}
\end{figure}

The structure of the eyewall is particularly evident in Figure~\ref{fig5}, 
which shows the distribution of azimuthal vorticity, $\omega_\phi/r$. 
It is clear that there are intense levels of 
azimuthal vorticity in the vicinity of the eyewall, and indeed it is natural 
to define the eyewall as the conical annulus of strong negative azimuthal 
vorticity. The eyewall then separates the eye from the primary vortex. 
Note also that an intense region of negative azimuthal vorticity has built 
up in the lower boundary layer, and it is shown in ODD17 that this is the 
ultimate source of the eyewall vorticity. A region of strong 
positive azimuthal vorticity is also evident between the lower boundary and the 
eyewall. As noted in ODD17, this is a local effect caused by 
the source term 
$\bfnabla \cdot \left[\left(\Gamma^2/r^4\right) \bfe_z\right]$
in (\ref{eqomegaphi}), 
which is particularly large near the base of the eyewall. However, since this source 
term takes the form of a flux it cannot contribute to the mean azimuthal vorticity in 
the eyewall (see ODD17). 

The general structure of the flow shown in Figures~\ref{fig3}, \ref{fig4} and 
\ref{fig5} is typical of all of our simulations which exhibit an eye. 
However, the scaling of the various velocity
components and the characteristic thickness of the bottom boundary layer depends on
the precise values of the control parameters. Let us start with some observations about
the thickness of the bottom boundary layer.

\begin{figure}
\centerline{
\includegraphics[width=0.8\textwidth]{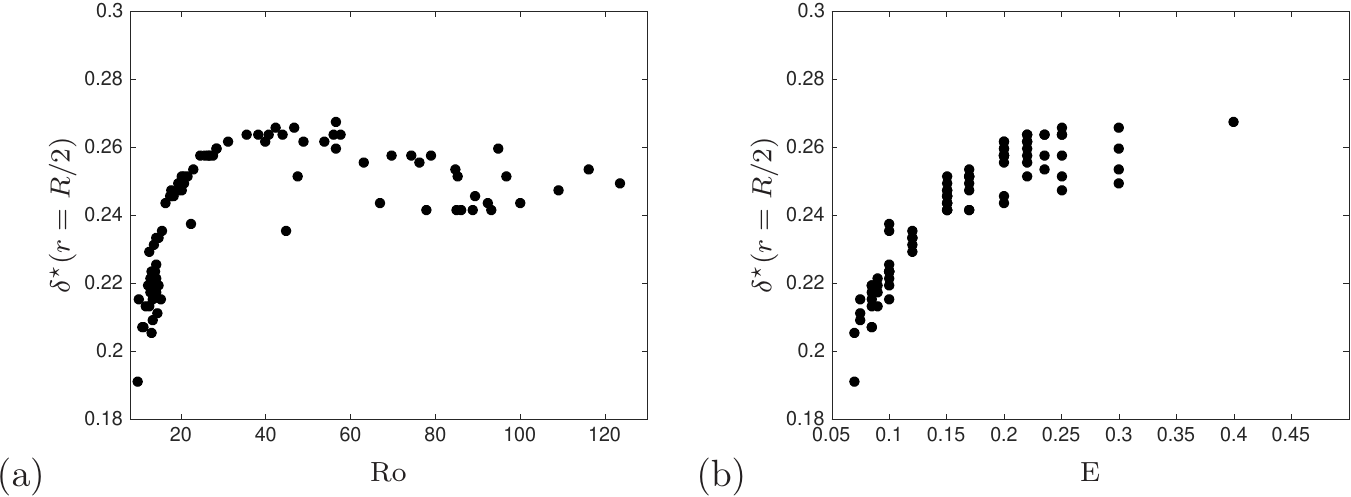}
}
\caption{The dimensionless boundary-layer thickness at mid radius, 
$\delta^\star(r=R/2)$, as a function of: (a) $\Ro$, and (b) $\Ek$.}
\label{fig6}
\end{figure}

\begin{figure}
\centerline{
\includegraphics[width=0.8\textwidth]{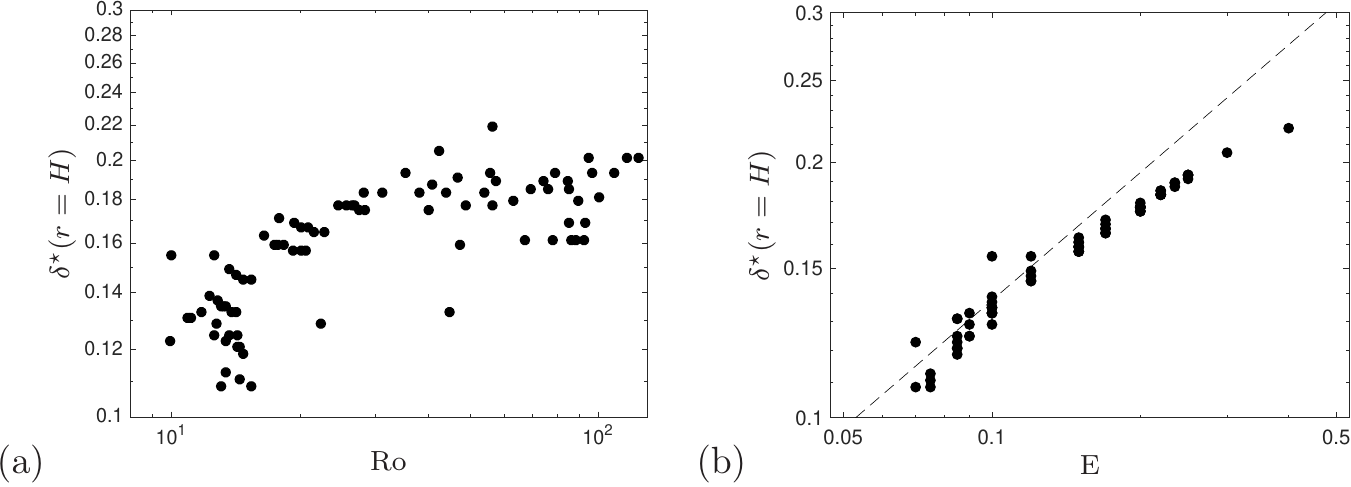}
}
\caption{The dimensionless boundary-layer thickness at $r = H$, $\delta^\star(r=H)$, 
as a function of: (a) $\Ro$, and (b) $\Ek$. The dashed line 
corresponds to $\delta^\star \sim \Ek^{1/2}$.}
\label{fig7}
\end{figure}

Figure~\ref{fig6} shows $\delta^\star=\delta/H$, the dimensionless boundary-layer 
thickness, evaluated at mid radius, $r = R/2$, and plotted as a function of $\Ro$ in 
Figure~\ref{fig6}(a) and $\Ek$ in Figure~\ref{fig6}(b). 
The results of all 87 numerical simulations 
for $\epsilon=0.1$ and $\Pr=0.1$ are shown. 
It is clear from Figure~\ref{fig6}(a) that there are two regimes. For $\Ro < 25$ 
we see that $\delta$ is an increasing function of $\Ro$, while for $\Ro > 30$ 
there is evidence that $\delta$ saturates at approximately $H/4$. 
We shall see shortly that these two distinct 
regimes also manifest themselves in the scaling laws for the velocity field, with a 
transition at around $\Ro \sim 25$. 
Figure~\ref{fig7} shows the same data, but for the location $r = H$. 
The boundary layer is now much thinner and there is some suggestion in 
Figure~\ref{fig7}(b) that $\delta^\star \sim \Ek^{1/2}$. This, in turn, 
suggests that the boundary layer at $r = H$ scales approximately 
as $\delta \sim (\nu/\Omega)^{1/2}$, as in a conventional Ekman layer.

\begin{figure}
\centerline{
\includegraphics[width=0.38\textwidth]{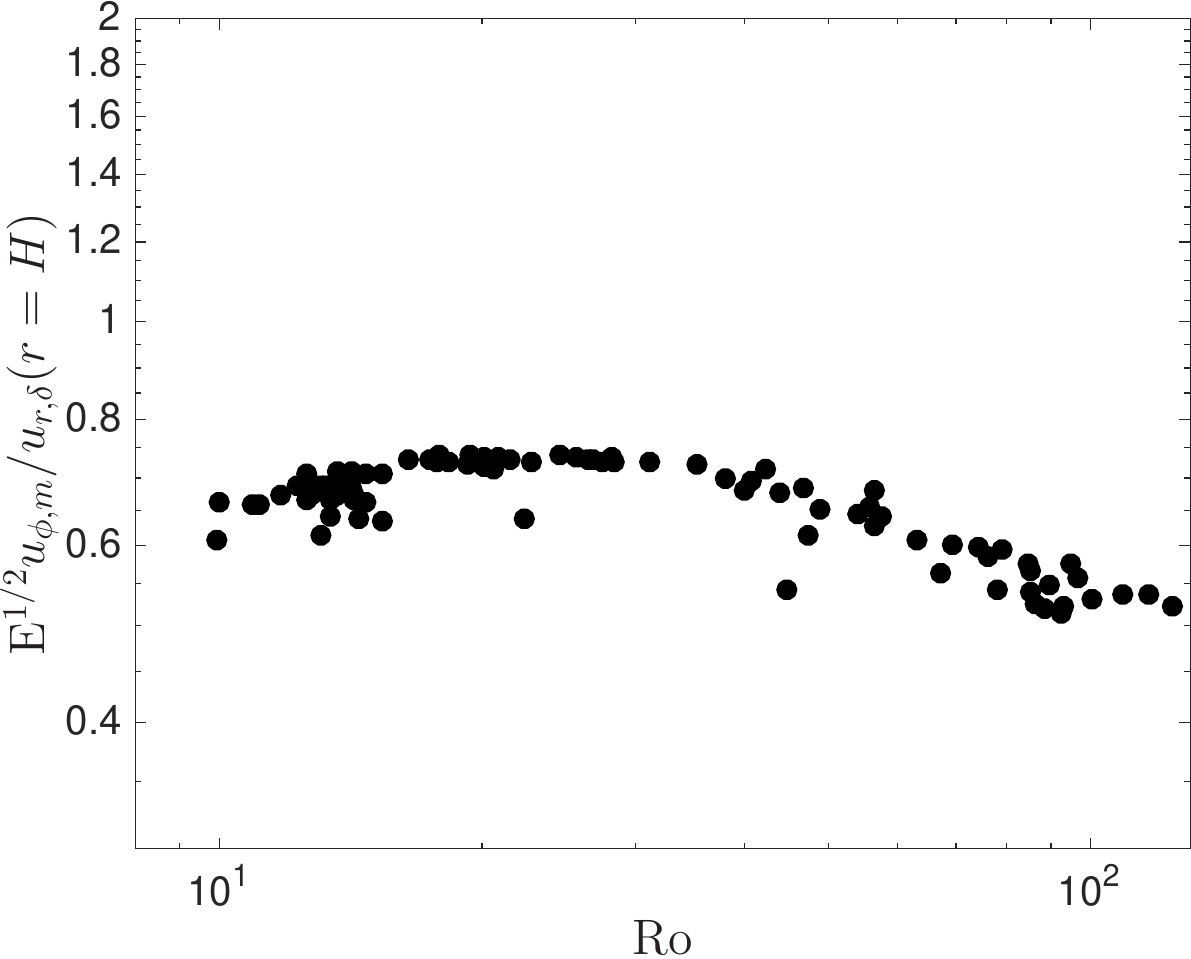}
}
\caption{The velocity ratio $\Ek^{1/2} u_{\phi,m}/u_{r,\delta}$ at $r=H$ as a 
function of $\Ro$.}
\label{fig8}
\end{figure}

We now consider the velocity ratio $u_{\phi,m}/u_{r,\delta}$.  
A preliminary analysis of the data indicates that this velocity 
ratio scales approximately as $u_{\phi,m}/u_{r,\delta} \sim \Ek^{-1/2}$, 
a scaling which is highlighted on Figure~\ref{fig8}, together 
with the remaining Rossby number dependence.

Regarding the scaling laws for the velocity field, it is instructive 
to integrate equation (\ref{equ}) once around a closed streamline. 
The inertial, pressure and Coriolis terms all drop out and we are left 
with the simple expression
\begin{equation}
\nu \oint \bfnabla^2 {\bf u} \cdot {\rm d} {\bf r} = \oint 
\left(\alpha \vartheta \bfg \right) \cdot {\rm d} {\bf r} \,.
\end{equation}
This represents an energy balance for a fluid particle as it is swept once around 
a closed streamline. In particular, it represents the balance between the viscous 
dissipation of energy and the work done on the fluid particle by the buoyancy force 
as the particle is carried around a streamline. For those cases in which the 
dissipation occurs primarily in the bottom boundary layer, this yields the estimate 
\begin{equation}
\nu \frac{u_{r,\delta}}{\delta^2} R \sim \alpha g \beta H^2 = V^2 \,.
\end{equation}
(We have taken advantage of the fact that $u_{\phi,m}/u_{r,\delta} \leq 1$ at 
most radii to omit the contribution from $u_\phi$.) If, in addition, we adopt the 
suggestion of Figure~\ref{fig7}(b) that the boundary layer thickness scales 
as $\delta \sim (\nu/\Omega)^{1/2}$, as in an Ekman layer, then we conclude that 
\begin{equation}
\frac{u_{r,\delta}}{V} \sim  \frac{V}{\Omega R}=\epsilon \Ro \,.
\label{scaling2}
\end{equation}

However, this estimate holds only when there is a well-developed boundary layer on 
the lower surface in which $\delta$ is much thinner than $H$. If the boundary layer 
grows to be of order $H$, on the other hand, the dissipation will be distributed 
throughout the bulk of the fluid and we would expect a different scaling law to hold. 
Figure~\ref{fig6}(a) tentatively suggests that scaling (\ref{scaling2}) might be 
appropriate for $\Ro < 25$, but not for $\Ro > 30$.
	
\begin{figure}
\centerline{
\includegraphics[width=0.8\textwidth]{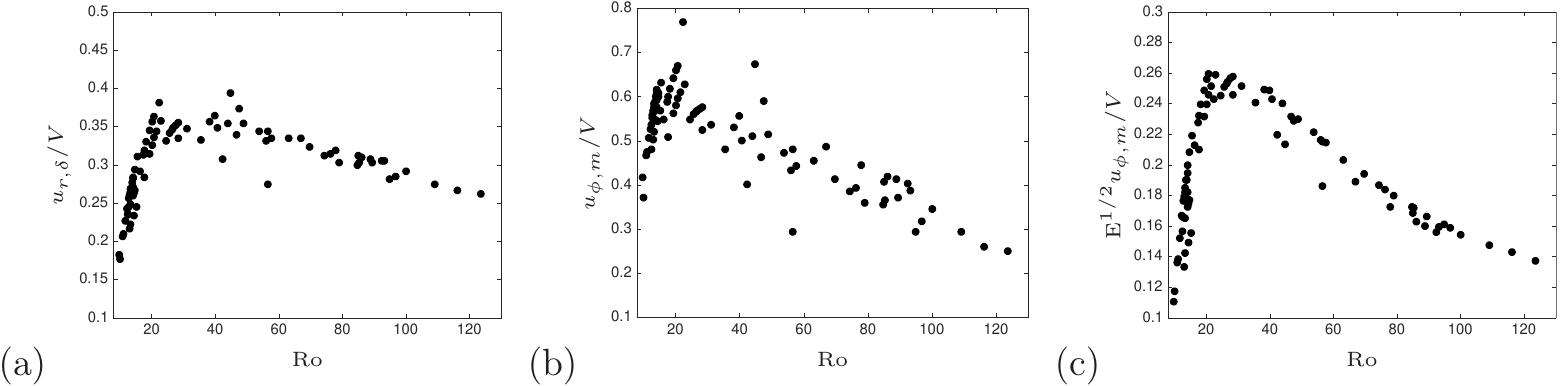}
}
\caption{
(a) $u_{r,\delta}/V$ at  $r = H$ as a function of $\Ro$. 
(b) $u_{\phi,m}/V$ versus $\Ro$.  
(c) $\Ek^{1/2} u_{\phi,m}/V$ versus $\Ro$.}
\label{fig9}
\end{figure}

Figure~\ref{fig9} shows: (a) $u_{r,\delta}/V$, (b) $u_{\phi,m}/V$, and (c) 
$\Ek^{1/2}u_{\phi,m}/V$, all evaluated at $r = H$ and plotted against $\Ro$. 
The data in Figure~\ref{fig9}(a) supports the idea that there are two regimes, 
with a transition at around $\Ro \sim 25$. 
Moreover, for $\Ro < 25$ there is some evidence in support of (\ref{scaling2}), 
while for $\Ro > 25$ the radial velocity saturates at $u_{r,\delta} \sim V$. 
There is considerably more scatter in Figure~\ref{fig9}(b), which shows 
$u_{\phi,m}/V$ as a function of $\Ro$. 
However, we have already noted that $u_{\phi,m}/u_{r,\delta} \sim \Ek^{-1/2}$  
at $r = H$ and so Figure~\ref{fig9}(c) shows the same data in the 
form $\Ek^{1/2} u_{\phi,m}/V$. 
The data is now reasonably well collapsed and again there is clear evidence of 
a transition in regimes at around $\Ro \sim 25$.

\begin{figure}
\centerline{
\includegraphics[width=0.8\textwidth]{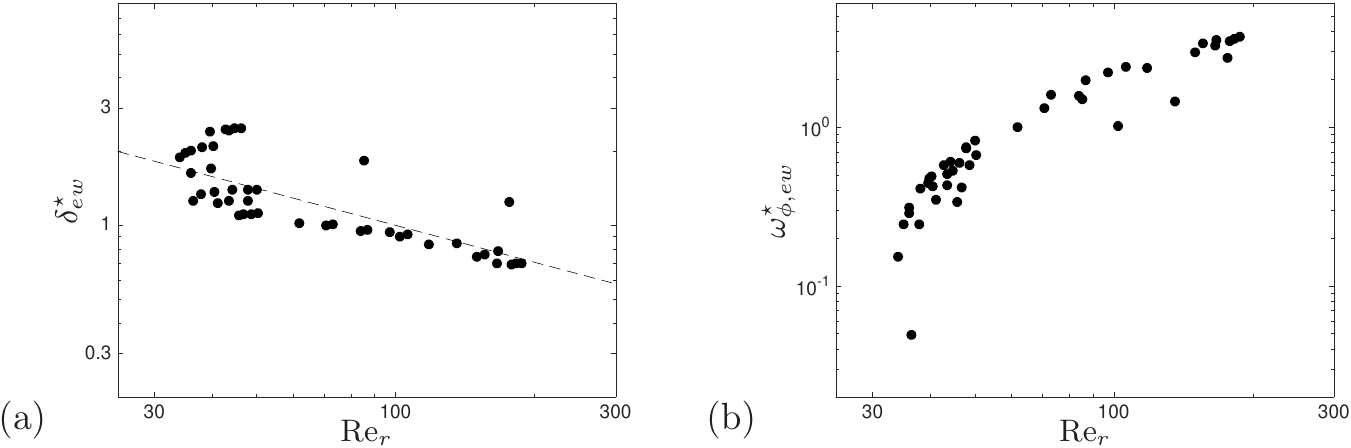}
}
\caption{Variation of (a) the width of the eyewall as
    measured by $\delta^\star_{ew}$ 
(the dashed line indicates a $\Ree_r^{-1/2}$ scaling)
and (b) the peak azimuthal vorticity in the eyewall 
$\omega^\star_{\phi , ew} \, ,$
both as a function of the Reynolds number $\Ree_r \, .$}
\label{fig10}
\end{figure}

Let us finally consider the thickness and strength of
the vorticity in the eyewall. 
We define the width of the eyewall $\delta^\star_{ew}$ as the horizontal
extent of the negative azimuthal vorticity at the height of the eye center $z_{eye}$
(we restrict ourselves here to cases in which an eye was observed).
It is to be expected that the width of the eyewall 
$\delta^\star_{ew}$ depends on the ratio of advection along the eyewall to
that of cross-stream diffusion. This appears to be well supported by
Figure~\ref{fig10}(a). Indeed the width of the eyewall appears to scale as
$\Ree_r ^{-1/2}$ as anticipated from a balance of streamwise advection and
cross-stream diffusion. The strength of
the vorticity in the eyewall can be estimated as $\omega^\star_{\phi , ew}
= - {\rm min}(\omega^\star_{\phi}(r,z_{eye})) \, .$ Figure~\ref{fig10}(b) shows
the increase of $\omega^\star_{\phi , ew}$ 
as the Reynolds number $\Ree_r$ is increased. The nature
of the supercritical bifurcation to an eye will be the object of the next section.

\section{The Transition to an Eye} 

It was noted in ODD17 that, for the limited set of cases examined, an eye would 
not form when $\Ree_r<37$. 
The reason is that the flow is then too diffusive for the boundary 
layer vorticity to be advected up in the bulk for the flow, and without this boundary 
layer vorticity, an eyewall cannot form. We now revisit this transition 
from no eye to an eye, focussing exclusively on the flow in the region 
$r \leq H$. We shall use as a 
measure of the strength of the eye the magnitude of the maximum downward velocity 
on the axis, $| \uu_z |^{\rm max}_{\rm r=0}$. The value of 
$| \uu_z |^{\rm max}_{\rm r=0}$ observed in each simulation is tabulated in 
the Appendix, with a zero entry for $| \uu_z |^{\rm max}_{\rm r=0}$ 
in the tables indicating that no eye formed in that simulation. 

\begin{figure}
\centerline{
\includegraphics[width=0.8\textwidth]{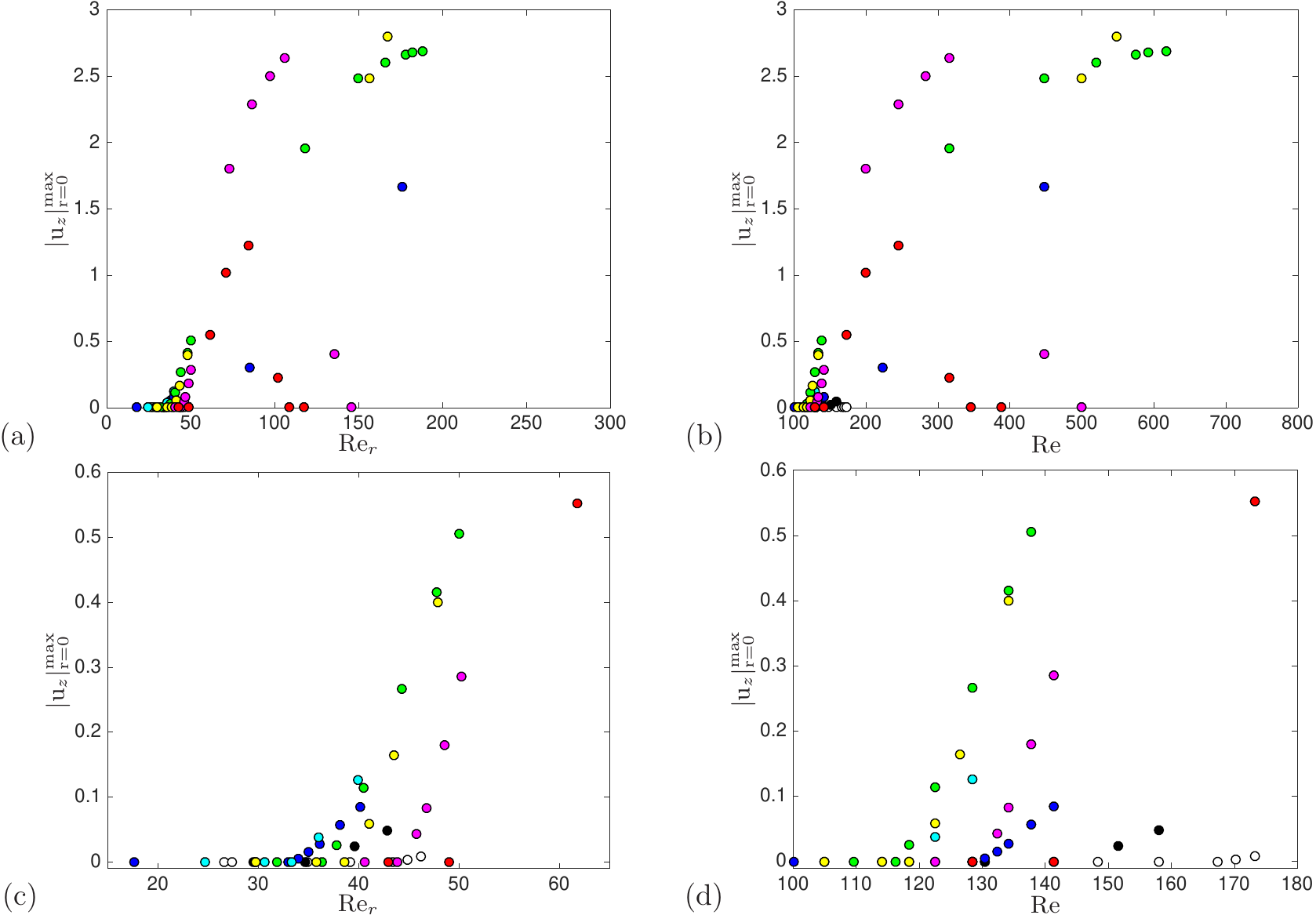}
}
\caption{Bifurcation diagrams of $| \uu_z |^{\rm max}_{\rm r=0}$ versus 
Reynolds number, with the bottom panels focused on 
the region of the bifurcation. 
The Ekman number is allowed to vary, but $\Pr$ and $\epsilon$ are fixed at 
$\Pr = \epsilon = 0.1$. 
The colour code indicates the values of $\Ek$, with 
$\Ek = 0.085$ (white), 
$\Ek = 0.090$ (black), 
$\Ek = 0.10$ (blue), 
$\Ek = 0.12$ (light blue),
$\Ek = 0.15$ (green),
$\Ek = 0.17$ (yellow),
$\Ek = 0.20$ (pink),
$\Ek = 0.22$ (red). 
(a), (c) $| \uu_z |^{\rm max}_{\rm r=0}$ plotted as a function of $\Ree_r$. 
(b), (d) $| \uu_z |^{\rm max}_{\rm r=0}$ plotted as a function of $\Ree$.}
\label{fig11}
\end{figure}

Figure~\ref{fig11} shows $| \uu_z |^{\rm max}_{\rm r=0}$ plotted as a 
function of (a) the measured $\Ree_r$ and  
(b) the controlled $\Ree$ for different values of $\Ek$. 
(Both $\Pr$ and $\epsilon$ are held fixed at $\epsilon=0.1$ and $\Pr=0.1$.) 
There is indeed a supercritical 
bifurcation to an eye at around $\Ree_{r,crit} \sim 40$, 
but there is also clear evidence that the critical 
Reynolds number, $\Ree_{r,crit}$, depends on $\Ek$, with $\Ree_{r,crit}$ varying 
from around $34$ up to a maximum of $50$. 
Considering the control parameter $\Ree$ yields
somewhat more scatter in the plot of $| \uu_z |^{\rm max}_{\rm r=0}$ versus $\Ree$, 
but the general trend is similar, with a supercritical bifurcation in the range 
$110 < \Ree_{crit} < 170$.

\begin{figure}
\centerline{
\includegraphics[width=0.8\textwidth]{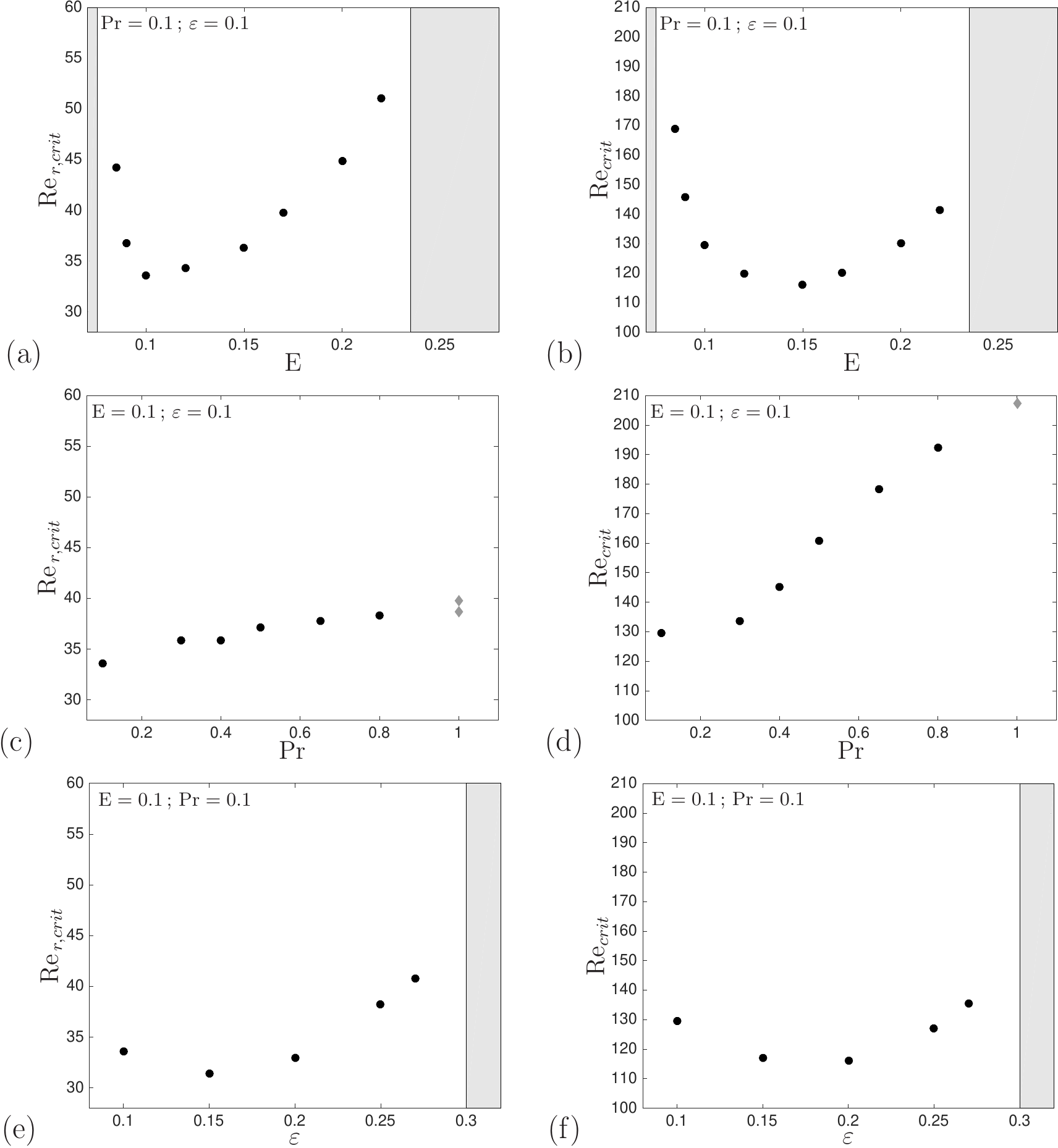}
}
\caption{
Critical Reynolds numbers $\Ree_{r,crit}$ (left column) 
and $\Ree_{crit}$ (right column). 
The grey areas denote an absence of an eye. 
The diamonds correspond to a hysteretic case.}
\label{fig12}
\end{figure}

The degree to which the critical Reynolds numbers $\Ree_{r,crit}$ and 
$\Ree_{crit}$ vary with $\Ek$, $\Pr$ and $\epsilon$ is 
explored in Figure~\ref{fig12}. 
Panels (a) and (b) show the dependency on $\Ek$, (c) and (d) 
the dependency on $\Pr$, and (e) and (f) the dependency on $\epsilon$. 
Interestingly, there is 
an optimum Ekman number for eye formation in the sense that $\Ree_{r,crit}$ and 
$\Ree_{crit}$ both exhibit minima. 
This minimum is around $\Ek \sim 0.1$ for $\Ree_{r,crit}$ 
and $\Ek \sim 0.15$ for $\Ree_{crit}$. Note also that no eyes are 
observed when $\Ek$ falls below $0.07$ or rises above $0.25$, 
as indicated by the grey areas 
in panels (a) and (b). We shall return to this observation shortly. 
There is also an optimal value of $\epsilon$ for eye formation, at around 
$\epsilon \sim 0.15-0.2$, with a complete absence 
of eyes for $\epsilon>0.3$ 
(at least for the range of parameters considered here). This suggests 
that a low aspect ratio is important for eye formation in this particular 
model problem.

The dependency of $\Ree_{r,crit}$  and $\Ree_{crit}$ on $\Pr$ is more complicated. 
While $\Ree_{r,crit}$ is only weakly dependent on $\Pr$, $\Ree_{crit}$ displays a 
marked dependency on $\Pr$, with $\Ree_{crit}$ rising sharply as $\Pr$ is increased. 
However, since $\Ree_{r,crit}$ is evaluated near the eyewall, and $\Ree_{crit}$ is 
a global quantity, we interpret the left-hand panel as indicating that $\Pr$ 
plays little or no role in the local dynamics of eye formation. The apparent 
dependency on $\Pr$ in the right-hand panel is then a manifestation of the 
fact that the global flow structure, and hence the ratio $\Ree_{r}/\Ree$, 
is a function of $\Pr$.

\begin{figure}
\centerline{
\includegraphics[width=0.9\textwidth]{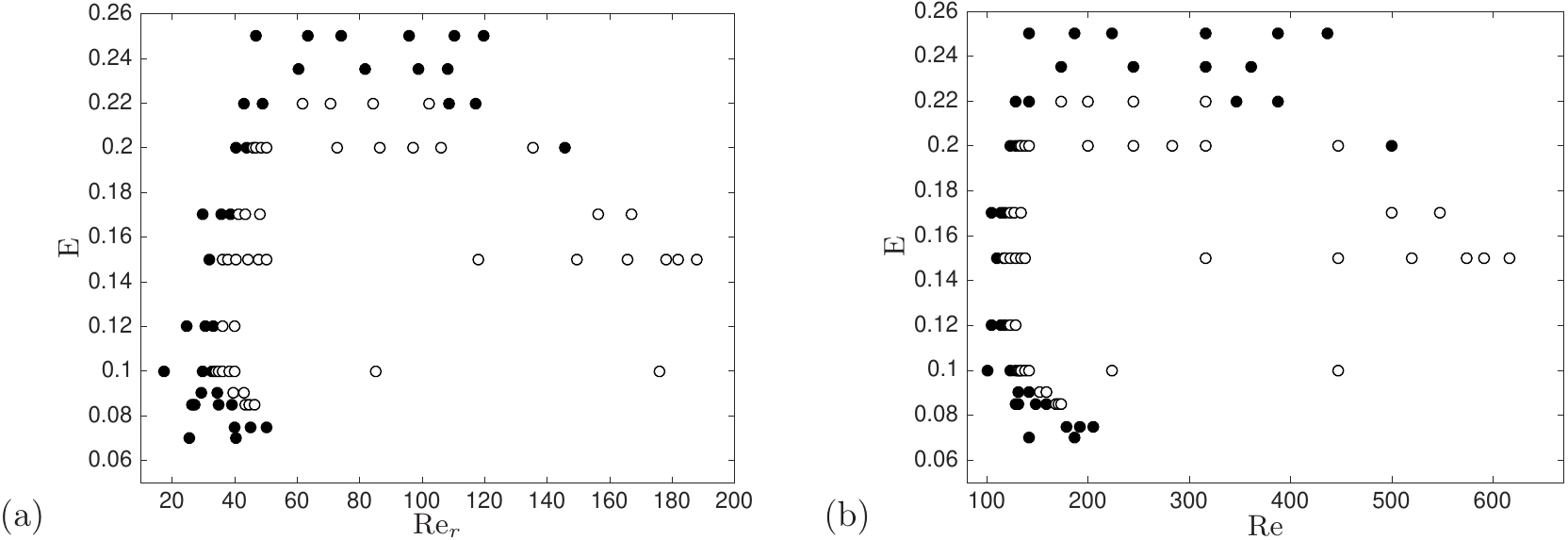}
}
\caption{Scatter plots of (a) $\Ek$ versus $\Ree_{r}$ and (b) $\Ek$ versus $\Ree$. 
The filled circles indicate the absence of an eye and the empty circles 
the presence of an eye. Both $\Pr$ and $\epsilon$ are held fixed at 
$\Pr =\epsilon = 0.1$.}
\label{fig13}
\end{figure}

While there are clearly lower bounds on $\Ree_{r,crit}$  and $\Ree_{crit}$ for eye 
formation, it is natural to ask if other conditions need to be satisfied. 
For example, the absence of eyes in Figure~\ref{fig12} for  $\Ek < 0.07$  
and  $\Ek > 0.25$ is intriguing. This is explored in Figure~\ref{fig13}, which 
presents scatter plots (or phase diagrams) of (a) $\Ek$ versus $\Ree_{r}$ 
and (b) $\Ek$ versus $\Ree$. In both cases the filled circles indicate the 
absence of an eye and the empty circles the presence of an eye. As in 
Figure~\ref{fig11}, $\Pr$ and $\epsilon$ are both held fixed at $\Pr =\epsilon = 0.1$. 
A more complex picture now emerges, with both upper and lower limits 
on $\Ek$ for eye formation, in addition to the lower bounds on $\Ree_{r,crit}$ 
and $\Ree_{crit}$.

The upper limit on $\Ek$ is to be expected from a consideration of global dynamics. 
That is to say, the presence of an eye rests on the formation of an eyewall, 
and this, in turn, requires the presence of a thin Ekman-like boundary layer 
surrounding the axis within which the fluid spirals inward. If $\Ek$ is too large,
then the Coriolis force acting in the bulk is unable to establish such a thin 
boundary layer in the face of strong viscous forces.

\section{Discussion}

\begin{figure}
\centerline{
\includegraphics[width=0.6\textwidth]{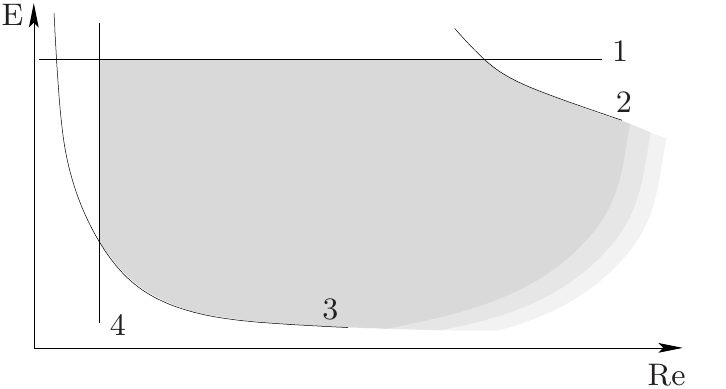}
}
\caption{Schematic structure of the phase diagram for the appearance of eyes in the 
model system considered in this paper. 
Eyes are not expected to form outside the shaded
region limited by the four curves shown. Extension of this domain to the 
right of this figure is so far unexplored.}
\label{fig14}
\end{figure}

Let us now pull together the results of the section~IV and 
summarise the conditions 
under which an eye is likely to form, at least in the particular model system 
investigated here. This is summarised in cartoon fashion in Figure~\ref{fig14} as a 
phase diagram of $\Ek$ versus $\Ree$, with $\Pr$ and $\epsilon$ both held fixed.

We suggest that the regime in which eyes are expected to form is 
limited by four 
curves. Line $4$ is the lower bound on $\Ree$ identified by ODD17, 
while line $1$ 
is the upper bound on $\Ek$ discussed above. Curves $2$ and $3$ are both of 
the form 
\begin{equation}
\Ek=\Ro_{crit}/\Ree\,,
\label{hyp}
\end{equation}
and represent upper and lower bounds on the Rossby number. While there is clear 
evidence in favour of lines $1$ and $4$ in Figure~\ref{fig13}, there is only 
moderate support for lines $2$ and $3$. 
However, upper and lower bounds on $\Ro$, 
as expressed by (\ref{hyp}), are conceptually necessary, as we now discuss.

The idea behind an upper bound on $\Ro$ is the assertion that the Coriolis force 
is essential for shaping the global flow pattern into a configuration favourable 
to eye formation. In particular, an appreciable Coriolis force acting on the bulk 
of the vortex is required to induce an Ekman-like boundary layer on the lower 
surface, without which an eyewall cannot form. So the Coriolis force cannot 
be significantly smaller than either the viscous or the inertial forces 
in the main body of the vortex. 
The restriction that the Coriolis force out-ways the viscous stresses leads to 
line $1$, as discussed above, while the requirement that it is at least as large 
as the inertial forces places an upper bound on $\Ro$ and yields line $2$. 

The lower bound on $\Ro$ stems from the fact that the local dynamics of eye formation 
occurs without any significant local influence from the buoyancy or Coriolis forces, 
as emphasised in ODD17. Indeed, it is essential that $\Ro_{\phi}$ is large 
(or at least greater than unity) at $r = H$, as otherwise quasi-geostrophy near 
the axis would prevent the lower boundary separating to form a conical shear 
layer, and hence prevent the formation of an eyewall. So we require 
$\Ro_{\phi}>1$ for 
an eye to form and this, in turn, suggests a lower bound on $\Ro$ in the bulk of 
the flow, thus leading to curve $3$. 
Certainly, it is noticeable that in tropical cyclones $\Ro_{\phi}$ near the eyewall 
is invariably significantly larger than unity.

In summary, then, there is clear supporting evidence for lines $1$ and $4$ 
in Figure~\ref{fig13}, and also some support for lines $2$ 
and $3$. Never-the-less, conceptually consistency requires an upper bound 
on $\Ro$ and a lower bound on $\Ro_{\phi}$ for eye formation, at least 
for the particular model system considered here.

It is interesting to consider 
the applicability of our simplified 
model to large-scale cyclonic vortices occurring in atmospheric flows, 
such as tropical cyclones. Of course, one must be 
cautious in such attempts, and
it is important to stress that certain essential characteristics of atmospheric 
vortices have been dropped in the present model. These include vertical 
stratification, spatially varying and anisotropic eddy viscosity, as well
as latent heat release due to water vapour condensation. 
However, most large-scale atmospheric vortices (tropical 
cyclones, medicanes, polar lows) exhibit an eye, which may be related to 
the eye in our simplified model.

The appropriate level of 
turbulent diffusion required to model a tropical cyclone is a poorly
constrained quantity \citep[]{Rotunno09}. It is most certainly non-uniform in
space and anisotropic. For the sake of simplicity, we could estimate an order
of magnitude for the relevant Ekman number, based on eddy viscosities in the
range $1 \rightarrow 10^3$~m$^2$.s$^{-1}$ and latitudes varying between some 
$10 ^\circ$
and $30 ^\circ$. This yields an estimate of the Ekman number in the range 
$10^{-4} \rightarrow 0.2 \, .$
Dropwindsonde observations in actual tropical cyclones indicate that the
inward radial flow above the boundary layer and at a radius close to the
eye is smaller by a factor about 10 than the azimuthal flow at the same
location \citep[]{Giammanco13}. This suggests, via the ratio
$u_{\phi,m}/u_{r,\delta} \sim \Ek^{-1/2}\, ,$ an effective Ekman number of
the order of $10^{-2}$.
Such estimates happen to be consistent with estimates of the
eddy viscosity above the boundary layer as well as with observations of the
boundary layer thickness in actual tropical cyclones \citep[]{Bryan17}.
Reynolds and Rossby number estimates may then be constructed on the basis of
such eddy viscosity orders of magnitude and in situ measurements 
of the inward radial flow (typically $5$~m.s$^{-1}$) and azimuthal flow
(typically $50$~m.s$^{-1}$). These estimates suggest $\Ree_r$ 
lies in the range $10^2 \rightarrow 10^5$
and $\Ro_\phi$ in the range $70 \rightarrow 220$, 
which include the parameter range covered by
our numerical study.
Another encouraging observation concerns the tilt of the eyewall. Airborne
Doppler radar data indicate that the eyewall in hurricanes is on average
characterised by a tilt angle of some $45^\circ$ \citep[]{Hazelton13,Stern14}, 
comparable to the tilt produced in the simplified model (see Figure~\ref{fig5}).
These observations may indicate that the fluid mechanics model presented
here, albeit simplified, is not irrelevant to some aspects of the 
dynamics of tropical cyclones, and could capture
some of the important physical mechanisms.

We should stress however the important distinction between
the large-scale vortices discussed here and elongated 
atmospheric vortices, such as 
tornadoes or dust devils. These tornado-like vortices have an inversed 
aspect ratio compared to our model. They do exhibit an eye-like structure, 
possibly associated with vortex breakdown. However, this breakdown is 
characterised by a much steeper wall
\citep[see for example][]{Fiedler94, Nolan05, RotunnoRev}.
Such phenomena correspond to a different configuration (both in terms of
aspect ratio and of controlling parameters), and 
our model is not relevant to such flows. Rather, we intend to model 
atmospheric vortices characterised by a large horizontal scale.

\section{Conclusions}

We have extended the study of ODD17, establishing scaling laws for the flow and 
mapping out the conditions under which an eye will form. We have shown that, to 
leading order, the velocity scales on $V=(\alpha g \beta)^{1/2}H$, and that 
the two most important parameters controlling the dynamics are $\Ree={V H}/{\nu}$ and 
$\Ro={V}/\left(\Omega H \right)$, with $\Pr$ and $\epsilon$ playing an important but 
secondary role. We have also shown that the criterion for eye formation in ODD17 
is too simplistic, and that upper and lower bounds on $\Ro$, as well as an upper 
bound on $\Ek$, must be taken into consideration.

\clearpage

\appendix
\section{Numerical results}

The following table presents 
the controlling parameters $\Ray$, $\Ro$, $\Ree$, 
and the diagnostic quantities 
$\Ro_r$, $\Ree_r$ and $|\uu_z|^{\rm max}_{\rm r=0}$
in our numerical database.

\def\baselinestretch{0.3}
\renewcommand{\thetable}{I\alph{table}} 
\begin{table}
\caption{Numerical results.}
\centering
\begin{tabular}{rrrrrr}
\hline
$\Ray$\ \   & \hskip 1cm $\Ro$    & \hskip 1cm $\Ree$  
            & \hskip 1cm $\Ro_r$  & \hskip 1cm $\Ree_r$
            & \hskip 1cm $|\uu_z|^{\rm max}_{\rm r=0}$ \\
\hline
\multicolumn{6}{c}{$\epsilon = 0.1$ \ $\Ek = 0.07$ \ $\Pr=0.1$}\\
2000 & 9.90 & 141.42 & 1.80 & 25.72 & 0 \\
3500 & 13.10 & 187.08 & 2.84 & 40.56 & 0 \\[1mm]
\multicolumn{6}{c}{$\epsilon = 0.1$ \ $\Ek = 0.075$ \ $\Pr=0.1$}\\
3200 & 13.42 & 178.89 & 2.99 & 39.82 & 0\\
3700 & 14.43 & 192.35 & 3.38 & 45.00 & 0\\
4200 & 15.37 & 204.94 & 3.77 & 50.23 & 0\\[1mm]
\multicolumn{6}{c}{$\epsilon = 0.1$ \ $\Ek = 0.085$ \ $\Pr=0.1$}\\
1650 & 10.92 & 128.45 & 2.26 & 26.56 & 0\\
1700 & 11.08 & 130.38 & 2.33 & 27.36 & 0\\
2200 & 12.61 & 148.32 & 2.96 & 34.86 & 0\\
2500 & 13.44 & 158.11 & 3.33 & 39.19 & 0\\
2800 & 14.22 & 167.33 & 3.69 & 43.46 & 0\\
2900 & 14.47 & 170.29 & 3.81 & 44.87 & 0.0023\\
3000 & 14.72 & 173.21 & 3.93 & 46.25 & 0.0076\\[1mm]
\multicolumn{6}{c}{$\epsilon = 0.1$ \ $\Ek = 0.09$ \ $\Pr=0.1$}\\
1700 & 11.73 & 130.38 & 2.66 & 29.55 & 0\\
2000 & 12.73 & 141.42 & 3.12 & 34.69 & 0\\
2300 & 13.65 & 151.66 & 3.57 & 39.63 & 0.0230\\
2500 & 14.23 & 158.11 & 3.86 & 42.87 & 0.0486\\[1mm]
\multicolumn{6}{c}{$\epsilon = 0.1$ \ $\Ek = 0.1$ \ $\Pr=0.1$}\\
1000 & 10.00 & 100.00 & 1.77 & 17.66 & 0\\
1500 & 12.25 & 122.47 & 2.97 & 29.71 & 0\\
1650 & 12.85 & 128.45 & 3.30 & 32.96 & 0\\
1700 & 13.04 & 130.38 & 3.40 & 34.01 & 0.0041\\
1750 & 13.23 & 132.29 & 3.51 & 35.06 & 0.0150\\
1800 & 13.42 & 134.16 & 3.61 & 36.10 & 0.0281\\
1900 & 13.78 & 137.84 & 3.81 & 38.14 & 0.0563\\
2000 & 14.14 & 141.42 & 4.01 & 40.15 & 0.0840\\
5000 & 22.36 & 223.61 & 8.53 & 85.28 & 0.3024\\
20000 & 44.72 & 447.21 & 17.61 & 176.10 & 1.6691\\[1mm]
\multicolumn{6}{c}{$\epsilon = 0.1$ \ $\Ek = 0.1$ \ $\Pr=0.3$}\\
2000 & 8.16 & 81.65 & 1.90 & 18.95 & 0\\
4000 & 11.55 & 115.47 & 3.03 & 30.33 & 0\\
5000 & 12.91 & 129.10 & 3.46 & 34.56 & 0\\
5500 & 13.54 & 135.40 & 3.64 & 36.44 & 0.0168\\
6000 & 14.14 & 141.42 & 3.82 & 38.20 & 0.0699\\
8000 & 16.33 & 163.30 & 4.43 & 44.26 & 0.3565\\
9000 & 17.32 & 173.21 & 4.68 & 46.83 & 0.4369\\[1mm]
\hline 
\end{tabular}
\label{tab:data}
\end{table}

\begin{table}
\caption{Numerical results (cont.)}
\centering
\begin{tabular}{rrrrrr}
\hline
$\Ray$\ \   & \hskip 1cm $\Ro$    & \hskip 1cm $\Ree$  
            & \hskip 1cm $\Ro_r$  & \hskip 1cm $\Ree_r$
            & \hskip 1cm $|\uu_z|^{\rm max}_{\rm r=0}$ \\
\hline
\multicolumn{6}{c}{$\epsilon = 0.1$ \ $\Ek = 0.1$ \ $\Pr=0.4$}\\
6000 & 12.25 & 122.47 & 2.98 & 29.82 & 0\\
8000 & 14.14 & 141.42 & 3.50 & 34.96 & 0\\
8300 & 14.40 & 144.05 & 3.57 & 35.66 & 0\\
8450 & 14.53 & 145.34 & 3.60 & 35.99 & 0.0016\\
8600 & 14.66 & 146.63 & 3.63 & 36.33 & 0.0080\\
8800 & 14.83 & 148.32 & 3.68 & 36.78 & 0.0203\\
9000 & 15.00 & 150.00 & 3.72 & 37.22 & 0.0359\\[1mm]
\multicolumn{6}{c}{$\epsilon = 0.1$ \ $\Ek = 0.1$ \ $\Pr=0.5$}\\
6000 & 10.95 & 109.54 & 2.42 & 24.24 & 0\\
8000 & 12.65 & 126.49 & 2.86 & 28.58 & 0\\
10000 & 14.14 & 141.42 & 3.23 & 32.34 & 0\\
11500 & 15.17 & 151.66 & 3.49 & 34.88 & 0\\
12000 & 15.49 & 154.92 & 3.57 & 35.68 & 0\\
13000 & 16.12 & 161.25 & 3.72 & 37.21 & 0.0246\\
15000 & 17.32 & 173.21 & 3.99 & 39.85 & 0.6992\\[1mm]
\multicolumn{6}{c}{$\epsilon = 0.1$ \ $\Ek = 0.1$ \ $\Pr=0.65$}\\
16000 & 15.69 & 156.89 & 3.29 & 32.90 & 0\\
18000 & 16.64 & 166.41 & 3.51 & 35.09 & 0\\
19000 & 17.10 & 170.97 & 3.61 & 36.13 & 0\\
21000 & 17.97 & 179.74 & 3.81 & 38.11 & 0.0523\\
21250 & 18.08 & 180.81 & 3.83 & 38.34 & 0.0944\\[1mm]
\multicolumn{6}{c}{$\epsilon = 0.1$ \ $\Ek = 0.1$ \ $\Pr=0.8$}\\
8000 & 10.00 & 100.00 & 1.82 & 18.15 & 0\\
15000 & 13.69 & 136.93 & 2.61 & 26.08 & 0\\
25000 & 17.68 & 176.78 & 3.48 & 34.80 & 0\\
27000 & 18.37 & 183.71 & 3.63 & 36.35 & 0\\
28700 & 18.94 & 189.41 & 3.76 & 37.60 & 0\\
29300 & 19.14 & 191.38 & 3.80 & 38.04 & 0\\
29400 & 19.17 & 191.70 & 3.81 & 38.11 & 0\\
29500 & 19.20 & 192.03 & 3.82 & 38.18 & 0\\
29650 & 19.25 & 192.52 & 3.83 & 38.29 & 0.0005\\
29700 & 19.27 & 192.68 & 3.83 & 38.33 & 0.0020\\[1mm]
\multicolumn{6}{c}{$\epsilon = 0.1$ \ $\Ek = 0.1$ \ $\Pr=1$}\\
2000 & 4.47 & 44.72 & 0.62 & 6.22 & 0\\
40000 & 20.00 & 200.00 & 3.71 & 37.10 & 0\\
42000 & 20.49 & 204.94 & 3.81 & 38.15 & 0\\
43000 & 20.74 & 207.36 & 3.87 & 38.69 & 0\\
43000 & 20.74 & 207.36 & 3.98 & 39.76 & 3.1408\\
44000 & 20.98 & 209.76 & 4.03 & 40.34 & 3.2326\\
45000 & 21.21 & 212.13 & 4.09 & 40.90 & 3.3116\\[1mm]
\multicolumn{6}{c}{$\epsilon = 0.1$ \ $\Ek = 0.12$ \ $\Pr=0.1$}\\
1100 & 12.59 & 104.88 & 2.97 & 24.72 & 0\\
1300 & 13.68 & 114.02 & 3.67 & 30.59 & 0\\
1400 & 14.20 & 118.32 & 4.00 & 33.36 & 0\\
1500 & 14.70 & 122.47 & 4.32 & 36.03 & 0.0384\\
1650 & 15.41 & 128.45 & 4.79 & 39.90 & 0.1263\\[1mm]
\hline 
\end{tabular}
\label{tab:data2}
\end{table}

\begin{table}
\caption{Numerical results (cont.)}
\centering
\begin{tabular}{rrrrrr}
\hline
$\Ray$\ \   & \hskip 1cm $\Ro$    & \hskip 1cm $\Ree$  
            & \hskip 1cm $\Ro_r$  & \hskip 1cm $\Ree_r$
            & \hskip 1cm $|\uu_z|^{\rm max}_{\rm r=0}$ \\
\hline
\multicolumn{6}{c}{$\epsilon = 0.1$ \ $\Ek = 0.15$ \ $\Pr=0.1$}\\
1200 & 16.43 & 109.54 & 4.78 & 31.90 & 0\\
1350 & 17.43 & 116.19 & 5.45 & 36.35 & 0.0003\\
1400 & 17.75 & 118.32 & 5.67 & 37.77 & 0.0260\\
1500 & 18.37 & 122.47 & 6.07 & 40.49 & 0.1134\\
1650 & 19.27 & 128.45 & 6.64 & 44.29 & 0.2669\\
1800 & 20.12 & 134.16 & 7.18 & 47.83 & 0.4154\\
1900 & 20.68 & 137.84 & 7.51 & 50.05 & 0.5049\\
10000 & 47.43 & 316.23 & 17.70 & 118.03 & 1.9520\\
20000 & 67.08 & 447.21 & 22.44 & 149.59 & 2.4840\\
27000 & 77.94 & 519.62 & 24.85 & 165.68 & 2.6067\\
33000 & 86.17 & 574.46 & 26.71 & 178.05 & 2.6655\\
35000 & 88.74 & 591.61 & 27.30 & 182.00 & 2.6783\\
38000 & 92.47 & 616.44 & 28.17 & 187.78 & 2.6902\\[1mm]
\multicolumn{6}{c}{$\epsilon = 0.1$ \ $\Ek = 0.17$ \ $\Pr=0.1$}\\
1100 & 17.83 & 104.88 & 5.06 & 29.79 & 0\\
1300 & 19.38 & 114.02 & 6.09 & 35.84 & 0\\
1400 & 20.11 & 118.32 & 6.55 & 38.55 & 0\\
1500 & 20.82 & 122.47 & 6.99 & 41.12 & 0.0583\\
1600 & 21.50 & 126.49 & 7.40 & 43.51 & 0.1636\\
1800 & 22.81 & 134.16 & 8.15 & 47.92 & 0.4005\\
25000 & 85.00 & 500.00 & 26.55 & 156.18 & 2.4819\\
30000 & 93.11 & 547.72 & 28.41 & 167.12 & 2.8008\\[1mm]
\multicolumn{6}{c}{$\epsilon = 0.1$ \ $\Ek = 0.2$ \ $\Pr=0.1$}\\
1500 & 24.49 & 122.47 & 8.13 & 40.63 & 0\\
1650 & 25.69 & 128.45 & 8.77 & 43.85 & 0\\
1750 & 26.46 & 132.29 & 9.16 & 45.82 & 0.0429\\
1800 & 26.83 & 134.16 & 9.35 & 46.77 & 0.0835\\
1900 & 27.57 & 137.84 & 9.71 & 48.55 & 0.1797\\
2000 & 28.28 & 141.42 & 10.05 & 50.25 & 0.2859\\
4000 & 40.00 & 200.00 & 14.58 & 72.92 & 1.8018\\
6000 & 48.99 & 244.95 & 17.32 & 86.62 & 2.2864\\
8000 & 56.57 & 282.84 & 19.42 & 97.08 & 2.5055\\
10000 & 63.25 & 316.23 & 21.17 & 105.86 & 2.6395\\
20000 & 89.44 & 447.21 & 27.10 & 135.50 & 0.4030\\
25000 & 100.00 & 500.00 & 29.13 & 145.65 & 0\\[1mm]
\multicolumn{6}{c}{$\epsilon = 0.1$ \ $\Ek = 0.22$ \ $\Pr=0.1$}\\
1650 & 28.26 & 128.45 & 9.45 & 42.96 & 0\\
2000 & 31.11 & 141.42 & 10.80 & 49.08 & 0\\
3000 & 38.11 & 173.21 & 13.59 & 61.77 & 0.5521\\
4000 & 44.00 & 200.00 & 15.58 & 70.83 & 1.0177\\
6000 & 53.89 & 244.95 & 18.51 & 84.15 & 1.2230\\
10000 & 69.57 & 316.23 & 22.47 & 102.12 & 0.2218\\
12000 & 76.21 & 346.41 & 23.93 & 108.80 & 0\\
15000 & 85.21 & 387.30 & 25.80 & 117.28 & 0\\[1mm]
\multicolumn{6}{c}{$\epsilon = 0.1$ \ $\Ek = 0.235$ \ $\Pr=0.1$}\\
3000 & 40.70 & 173.21 & 14.20 & 60.41 & 0\\
6000 & 57.56 & 244.95 & 19.25 & 81.92 & 0\\
10000 & 74.31 & 316.23 & 23.21 & 98.77 & 0\\
13000 & 84.73 & 360.56 & 25.41 & 108.11 & 0\\[1mm]
\hline 
\end{tabular}
\label{tab:data3}
\end{table}

\begin{table}
\caption{Numerical results (cont.)}
\centering
\begin{tabular}{rrrrrr}
\hline
$\Ray$\ \   & \hskip 1cm $\Ro$    & \hskip 1cm $\Ree$  
            & \hskip 1cm $\Ro_r$  & \hskip 1cm $\Ree_r$
            & \hskip 1cm $|\uu_z|^{\rm max}_{\rm r=0}$ \\
\hline
\multicolumn{6}{c}{$\epsilon = 0.1$ \ $\Ek = 0.25$ \ $\Pr=0.1$}\\
2000 & 35.36 & 141.42 & 11.76 & 47.03 & 0\\
3500 & 46.77 & 187.08 & 15.85 & 63.41 & 0\\
5000 & 55.90 & 223.61 & 18.49 & 73.97 & 0\\
10000 & 79.06 & 316.23 & 23.95 & 95.81 & 0\\
15000 & 96.82 & 387.30 & 27.59 & 110.36 & 0\\
19000 & 108.97 & 435.89 & 29.95 & 119.78 & 0\\[1mm]
\multicolumn{6}{c}{$\epsilon = 0.1$ \ $\Ek = 0.3$ \ $\Pr=0.1$}\\
2000 & 42.43 & 141.42 & 13.03 & 43.44 & 0\\
10000 & 94.87 & 316.23 & 26.64 & 88.81 & 0\\
15000 & 116.19 & 387.30 & 30.97 & 103.22 & 0\\
17000 & 123.69 & 412.31 & 32.43 & 108.10 & 0\\[1mm]
\multicolumn{6}{c}{$\epsilon = 0.1$ \ $\Ek = 0.4$ \ $\Pr=0.1$}\\
2000 & 56.57 & 141.42 & 15.52 & 38.79 & 0\\[1mm]
\multicolumn{6}{c}{$\epsilon = 0.15$ \ $\Ek = 0.1$ \ $\Pr=0.1$}\\
1300 & 11.40 & 114.02 & 2.93 & 29.28 & 0\\
1400 & 11.83 & 118.32 & 3.24 & 32.44 & 0.0175\\
1500 & 12.25 & 122.47 & 3.55 & 35.45 & 0.0711\\
1700 & 13.04 & 130.38 & 4.10 & 40.99 & 0.1898\\[1mm]
\multicolumn{6}{c}{$\epsilon = 0.2$ \ $\Ek = 0.1$ \ $\Pr=0.1$}\\
1000 & 10.00 & 100.00 & 2.12 & 21.25 & 0\\
1200 & 10.95 & 109.54 & 2.83 & 28.33 & 0\\
1450 & 12.04 & 120.42 & 3.55 & 35.46 & 0.0588\\
1500 & 12.25 & 122.47 & 3.67 & 36.68 & 0.0878\\
2000 & 14.14 & 141.42 & 4.65 & 46.53 & 0.3919\\
4000 & 20.00 & 200.00 & 6.73 & 67.28 & 1.0136\\
10000 & 31.62 & 316.23 & 9.63 & 96.27 & 1.3347\\
20000 & 44.72 & 447.21 & 12.65 & 126.50 & 1.6677\\[1mm]
\multicolumn{6}{c}{$\epsilon = 0.25$ \ $\Ek = 0.1$ \ $\Pr=0.1$}\\
1500 & 12.25 & 122.47 & 3.54 & 35.44 & 0\\
2000 & 14.14 & 141.42 & 4.34 & 43.41 & 0.1366\\
2500 & 15.81 & 158.11 & 4.94 & 49.35 & 0.2948\\
4000 & 20.00 & 200.00 & 6.22 & 62.23 & 0.5584\\[1mm]
\multicolumn{6}{c}{$\epsilon = 0.27$ \ $\Ek = 0.1$ \ $\Pr=0.1$}\\
1000 & 10.00 & 100.00 & 2.26 & 22.62 & 0\\
2000 & 14.14 & 141.42 & 4.29 & 42.88 & 0.0168\\
3000 & 17.32 & 173.21 & 5.36 & 53.60 & 0.1041\\
5000 & 22.36 & 223.61 & 6.79 & 67.85 & 0.0026\\
7000 & 26.46 & 264.58 & 7.79 & 77.94 & 0\\[1mm]
\multicolumn{6}{c}{$\epsilon = 0.3$ \ $\Ek = 0.1$ \ $\Pr=0.1$}\\
1000 & 10.00 & 100.00 & 2.17 & 21.70 & 0\\
2000 & 14.14 & 141.42 & 4.09 & 40.85 & 0\\
4000 & 20.00 & 200.00 & 5.77 & 57.67 & 0\\
6000 & 24.49 & 244.95 & 6.80 & 67.96 & 0\\
9000 & 30.00 & 300.00 & 7.90 & 79.01 & 0\\
12000 & 34.64 & 346.41 & 8.74 & 87.39 & 0\\[1mm]
\hline 
\end{tabular}
\label{tab:data4}
\end{table}


\bibliography{biblio}

\providecommand{\noopsort}[1]{}
\begin{thebibliography}{16}%
\makeatletter
\providecommand \@ifxundefined [1]{%
 \@ifx{#1\undefined}
}%
\providecommand \@ifnum [1]{%
 \ifnum #1\expandafter \@firstoftwo
 \else \expandafter \@secondoftwo
 \fi
}%
\providecommand \@ifx [1]{%
 \ifx #1\expandafter \@firstoftwo
 \else \expandafter \@secondoftwo
 \fi
}%
\providecommand \natexlab [1]{#1}%
\providecommand \enquote  [1]{``#1''}%
\providecommand \bibnamefont  [1]{#1}%
\providecommand \bibfnamefont [1]{#1}%
\providecommand \citenamefont [1]{#1}%
\providecommand \href@noop [0]{\@secondoftwo}%
\providecommand \href [0]{\begingroup \@sanitize@url \@href}%
\providecommand \@href[1]{\@@startlink{#1}\@@href}%
\providecommand \@@href[1]{\endgroup#1\@@endlink}%
\providecommand \@sanitize@url [0]{\catcode `\\12\catcode `\$12\catcode
  `\&12\catcode `\#12\catcode `\^12\catcode `\_12\catcode `\%12\relax}%
\providecommand \@@startlink[1]{}%
\providecommand \@@endlink[0]{}%
\providecommand \url  [0]{\begingroup\@sanitize@url \@url }%
\providecommand \@url [1]{\endgroup\@href {#1}{\urlprefix }}%
\providecommand \urlprefix  [0]{URL }%
\providecommand \Eprint [0]{\href }%
\providecommand \doibase [0]{http://dx.doi.org/}%
\providecommand \selectlanguage [0]{\@gobble}%
\providecommand \bibinfo  [0]{\@secondoftwo}%
\providecommand \bibfield  [0]{\@secondoftwo}%
\providecommand \translation [1]{[#1]}%
\providecommand \BibitemOpen [0]{}%
\providecommand \bibitemStop [0]{}%
\providecommand \bibitemNoStop [0]{.\EOS\space}%
\providecommand \EOS [0]{\spacefactor3000\relax}%
\providecommand \BibitemShut  [1]{\csname bibitem#1\endcsname}%
\let\auto@bib@innerbib\@empty
\bibitem [{\citenamefont {Oruba}\ \emph {et~al.}(2017)\citenamefont {Oruba},
  \citenamefont {Davidson},\ and\ \citenamefont {Dormy}}]{ODD17}%
  \BibitemOpen
  \bibfield  {author} {\bibinfo {author} {\bibfnamefont {L.}~\bibnamefont
  {Oruba}}, \bibinfo {author} {\bibfnamefont {P.~A.}\ \bibnamefont {Davidson}},
  \ and\ \bibinfo {author} {\bibfnamefont {E.}~\bibnamefont {Dormy}},\
  }\bibfield  {title} {\enquote {\bibinfo {title} {{Eye formation in rotating
  convection}},}\ }\href@noop {} {\bibfield  {journal} {\bibinfo  {journal} {J.
  Fluid Mech.}\ }\textbf {\bibinfo {volume} {812}},\ \bibinfo {pages}
  {890--904} (\bibinfo {year} {2017})}\BibitemShut {NoStop}%
\bibitem [{\citenamefont {Lugt}(1983)}]{Lugt}%
  \BibitemOpen
  \bibfield  {author} {\bibinfo {author} {\bibfnamefont {H.~J.}\ \bibnamefont
  {Lugt}},\ }\href@noop {} {\emph {\bibinfo {title} {Vortex flow in nature and
  technology}}}\ (\bibinfo  {publisher} {Wiley},\ \bibinfo {year}
  {1983})\BibitemShut {NoStop}%
\bibitem [{\citenamefont {Pearce}(2005a)}]{Pearce}%
  \BibitemOpen
  \bibfield  {author} {\bibinfo {author} {\bibfnamefont {R.}~\bibnamefont
  {Pearce}},\ }\bibfield  {title} {\enquote {\bibinfo {title} {Why must
  hurricanes have eyes?}}\ }\href@noop {} {\bibfield  {journal} {\bibinfo
  {journal} {Weather}\ }\textbf {\bibinfo {volume} {60}},\ \bibinfo {pages}
  {19--24} (\bibinfo {year} {2005a})}\BibitemShut {NoStop}%
\bibitem [{\citenamefont {Smith}(2005)}]{Smith}%
  \BibitemOpen
  \bibfield  {author} {\bibinfo {author} {\bibfnamefont {R.~K.}\ \bibnamefont
  {Smith}},\ }\bibfield  {title} {\enquote {\bibinfo {title} {``{W}hy must
  hurricanes have eyes?'' revisited},}\ }\href@noop {} {\bibfield  {journal}
  {\bibinfo  {journal} {Weather}\ }\textbf {\bibinfo {volume} {60}},\ \bibinfo
  {pages} {326--328} (\bibinfo {year} {2005})}\BibitemShut {NoStop}%
\bibitem [{\citenamefont {Pearce}(2005b)}]{Pearce2}%
  \BibitemOpen
  \bibfield  {author} {\bibinfo {author} {\bibfnamefont {R.}~\bibnamefont
  {Pearce}},\ }\bibfield  {title} {\enquote {\bibinfo {title} {Comments on
  ``{W}hy must hurricanes have eyes?'' revisited},}\ }\href@noop {} {\bibfield
  {journal} {\bibinfo  {journal} {Weather}\ }\textbf {\bibinfo {volume} {60}},\
  \bibinfo {pages} {329--330} (\bibinfo {year} {2005b})}\BibitemShut {NoStop}%
\bibitem [{\citenamefont {{Rotunno}}(2014)}]{Rotunno14}%
  \BibitemOpen
  \bibfield  {author} {\bibinfo {author} {\bibfnamefont {R.}~\bibnamefont
  {{Rotunno}}},\ }\bibfield  {title} {\enquote {\bibinfo {title} {{Secondary
  circulations in rotating-flow boundary layers}},}\ }\href@noop {} {\bibfield
  {journal} {\bibinfo  {journal} {Aust. Met. Oceanogr. J.}\ }\textbf {\bibinfo
  {volume} {64}},\ \bibinfo {pages} {27--35} (\bibinfo {year}
  {2014})}\BibitemShut {NoStop}%
\bibitem [{\citenamefont {Turner}(1973)}]{Turner}%
  \BibitemOpen
  \bibfield  {author} {\bibinfo {author} {\bibfnamefont {J.~S.}\ \bibnamefont
  {Turner}},\ }\href@noop {} {\emph {\bibinfo {title} {Buoyancy Effects in
  Fluids}}}\ (\bibinfo  {publisher} {Cambridge Univ. Press},\ \bibinfo {year}
  {1973})\BibitemShut {NoStop}%
\bibitem [{\citenamefont {Davidson}(2013)}]{Davidson}%
  \BibitemOpen
  \bibfield  {author} {\bibinfo {author} {\bibfnamefont {P.~A.}\ \bibnamefont
  {Davidson}},\ }\href@noop {} {\emph {\bibinfo {title} {Turbulence in
  Rotating, Stratified and Electrically Conducting Fluids}}}\ (\bibinfo
  {publisher} {Cambridge Univ. Press},\ \bibinfo {year} {2013})\BibitemShut
  {NoStop}%
\bibitem [{\citenamefont {Rotunno}\ \emph {et~al.}(2009)\citenamefont
  {Rotunno}, \citenamefont {Chen}, \citenamefont {Wang}, \citenamefont {Davis},
  \citenamefont {Dudhia},\ and\ \citenamefont {Holland}}]{Rotunno09}%
  \BibitemOpen
  \bibfield  {author} {\bibinfo {author} {\bibfnamefont {R.}~\bibnamefont
  {Rotunno}}, \bibinfo {author} {\bibfnamefont {Y.}~\bibnamefont {Chen}},
  \bibinfo {author} {\bibfnamefont {W.}~\bibnamefont {Wang}}, \bibinfo {author}
  {\bibfnamefont {C.}~\bibnamefont {Davis}}, \bibinfo {author} {\bibfnamefont
  {J.}~\bibnamefont {Dudhia}}, \ and\ \bibinfo {author} {\bibfnamefont {G.~J.}\
  \bibnamefont {Holland}},\ }\bibfield  {title} {\enquote {\bibinfo {title}
  {{Large-eddy simulation of an idealized tropical cyclone}},}\ }\href@noop {}
  {\bibfield  {journal} {\bibinfo  {journal} {Bull. Amer. Meteor. Soc.}\
  }\textbf {\bibinfo {volume} {90}},\ \bibinfo {pages} {1783–1788} (\bibinfo
  {year} {2009})}\BibitemShut {NoStop}%
\bibitem [{\citenamefont {Giammanco}\ \emph {et~al.}(2013)\citenamefont
  {Giammanco}, \citenamefont {Schroeder},\ and\ \citenamefont
  {Powell}}]{Giammanco13}%
  \BibitemOpen
  \bibfield  {author} {\bibinfo {author} {\bibfnamefont {I.~M.}\ \bibnamefont
  {Giammanco}}, \bibinfo {author} {\bibfnamefont {J.~L.}\ \bibnamefont
  {Schroeder}}, \ and\ \bibinfo {author} {\bibfnamefont {M.~D.}\ \bibnamefont
  {Powell}},\ }\bibfield  {title} {\enquote {\bibinfo {title} {{GPS
  dropwindsonde and WSR-88D observations of tropical cy- clone vertical wind
  profiles and their characteristics.}}}\ }\href@noop {} {\bibfield  {journal}
  {\bibinfo  {journal} {Wea. Forecasting}\ }\textbf {\bibinfo {volume} {28}},\
  \bibinfo {pages} {77--99} (\bibinfo {year} {2013})}\BibitemShut {NoStop}%
\bibitem [{\citenamefont {Bryan}\ \emph {et~al.}(2017)\citenamefont {Bryan},
  \citenamefont {Worsnop}, \citenamefont {Lundquist},\ and\ \citenamefont
  {Zhang}}]{Bryan17}%
  \BibitemOpen
  \bibfield  {author} {\bibinfo {author} {\bibfnamefont {G.~H.}\ \bibnamefont
  {Bryan}}, \bibinfo {author} {\bibfnamefont {R.~P.}\ \bibnamefont {Worsnop}},
  \bibinfo {author} {\bibfnamefont {J.~K.}\ \bibnamefont {Lundquist}}, \ and\
  \bibinfo {author} {\bibfnamefont {J.~A.}\ \bibnamefont {Zhang}},\ }\bibfield
  {title} {\enquote {\bibinfo {title} {{A simple method for simulating wind
  profiles in the boundary layer of tropical cyclones.}}}\ }\href@noop {}
  {\bibfield  {journal} {\bibinfo  {journal} {Boundary-Layer Meteorol.}\
  }\textbf {\bibinfo {volume} {162}},\ \bibinfo {pages} {475--502} (\bibinfo
  {year} {2017})}\BibitemShut {NoStop}%
\bibitem [{\citenamefont {Hazelton}\ and\ \citenamefont
  {Hart}(2013)}]{Hazelton13}%
  \BibitemOpen
  \bibfield  {author} {\bibinfo {author} {\bibfnamefont {A.~T.}\ \bibnamefont
  {Hazelton}}\ and\ \bibinfo {author} {\bibfnamefont {R.~E.}\ \bibnamefont
  {Hart}},\ }\bibfield  {title} {\enquote {\bibinfo {title} {{Hurricane eyewall
  slope as determined from airborne radar reflectivity data: Composites and
  case studies.}}}\ }\href@noop {} {\bibfield  {journal} {\bibinfo  {journal}
  {Wea. Forecasting}\ }\textbf {\bibinfo {volume} {28}},\ \bibinfo {pages}
  {368–386} (\bibinfo {year} {2013})}\BibitemShut {NoStop}%
\bibitem [{\citenamefont {Stern}\ \emph {et~al.}(2014)\citenamefont {Stern},
  \citenamefont {Brisbois},\ and\ \citenamefont {Nolan}}]{Stern14}%
  \BibitemOpen
  \bibfield  {author} {\bibinfo {author} {\bibfnamefont {D.~P.}\ \bibnamefont
  {Stern}}, \bibinfo {author} {\bibfnamefont {J.~R.}\ \bibnamefont {Brisbois}},
  \ and\ \bibinfo {author} {\bibfnamefont {D.~S.}\ \bibnamefont {Nolan}},\
  }\bibfield  {title} {\enquote {\bibinfo {title} {{An expanded dataset of
  hurricane eyewall sizes and slopes.}}}\ }\href@noop {} {\bibfield  {journal}
  {\bibinfo  {journal} {J. Atmos. Sci.}\ }\textbf {\bibinfo {volume} {71}},\
  \bibinfo {pages} {2747–2762} (\bibinfo {year} {2014})}\BibitemShut
  {NoStop}%
\bibitem [{\citenamefont {Fiedler}(1994)}]{Fiedler94}%
  \BibitemOpen
  \bibfield  {author} {\bibinfo {author} {\bibfnamefont {B.~H.}\ \bibnamefont
  {Fiedler}},\ }\bibfield  {title} {\enquote {\bibinfo {title} {{The
  thermodynamic speed limit and its violation in axisymmetric numerical
  simulations of tornado-like vortices.}}}\ }\href@noop {} {\bibfield
  {journal} {\bibinfo  {journal} {Atmos.–Ocean}\ }\textbf {\bibinfo {volume}
  {32}},\ \bibinfo {pages} {335–359} (\bibinfo {year} {1994})}\BibitemShut
  {NoStop}%
\bibitem [{\citenamefont {Nolan}(2005)}]{Nolan05}%
  \BibitemOpen
  \bibfield  {author} {\bibinfo {author} {\bibfnamefont {D.~S.}\ \bibnamefont
  {Nolan}},\ }\bibfield  {title} {\enquote {\bibinfo {title} {{A new scaling
  for tornado-like vortices.}}}\ }\href@noop {} {\bibfield  {journal} {\bibinfo
   {journal} {J. Atmos. Sci.}\ }\textbf {\bibinfo {volume} {62}},\ \bibinfo
  {pages} {2639–45} (\bibinfo {year} {2005})}\BibitemShut {NoStop}%
\bibitem [{\citenamefont {Rotunno}(2013)}]{RotunnoRev}%
  \BibitemOpen
  \bibfield  {author} {\bibinfo {author} {\bibfnamefont {R.}~\bibnamefont
  {Rotunno}},\ }\bibfield  {title} {\enquote {\bibinfo {title} {{The Fluid
  Dynamics of Tornadoes.}}}\ }\href@noop {} {\bibfield  {journal} {\bibinfo
  {journal} {Annu. Rev. Fluid Mech.}\ }\textbf {\bibinfo {volume} {45}},\
  \bibinfo {pages} {59--84} (\bibinfo {year} {2013})}\BibitemShut {NoStop}%
\end{thebibliography}%

\end{document}